\documentclass[9pt,twocolumn,twoside]{optica}
\setboolean{shortarticle}{false}
\setboolean{minireview}{false}

\usepackage{graphicx}
\usepackage{dcolumn}
\usepackage{bm}
 \usepackage{tensor}
 \usepackage{color}
\usepackage{times}
\usepackage{siunitx}
\usepackage{physics}

\title{Quantum-limited estimation of the axial separation of two incoherent point sources}

\author[1,*]{Yiyu Zhou}
\author[2]{Jing Yang}
\author[1]{Jeremy D. Hassett}
\author[3]{Seyed Mohammad Hashemi Rafsanjani}
\author[4]{Mohammad Mirhosseini}
\author[1,2,5]{A. Nick Vamivakas}
\author[2,6]{Andrew N. Jordan}
\author[7,$\dagger$]{Zhimin Shi}
\author[1,8,$\ddagger$]{Robert W. Boyd}

\affil[1]{The Institute of Optics, University of Rochester, Rochester, New York 14627, USA}
\affil[2]{Department of Physics and Astronomy, University of Rochester, Rochester, New York 14627, USA}
\affil[3]{Department of Physics, University of Miami, Coral Gables, Florida 33146, USA}
\affil[4]{Thomas J. Watson, Sr., Laboratory of Applied Physics, California Institute of Technology, Pasadena, California 91125, USA}
\affil[5]{Materials Science Program, University of Rochester, Rochester, New York 14627, USA}
\affil[6]{Institute for Quantum Studies, Chapman University, Orange, California 92866, USA}
\affil[7]{Department of Physics, University of South Florida, Tampa, Florida 33620, USA}
\affil[8]{Department of Physics, University of Ottawa, Ottawa, Ontario K1N 6N5, Canada}
\affil[*]{Corresponding author: yzhou62@ur.rochester.edu}
\affil[$\dagger$]{email: zhiminshi@usf.edu}
\affil[$\ddagger$]{email: boyd@optics.rochester.edu}

\ociscodes{(100.6640) Superresolution; (030.4070) Modes; (270.5585) Quantum information and processing.}

\begin{abstract}
Improving axial resolution is crucial for three-dimensional optical imaging systems. Here we present a scheme of axial superresolution for two incoherent point sources based on spatial mode demultiplexing. A radial mode sorter is used to losslessly decompose the optical fields into a radial mode basis set to extract the phase information associated with the axial positions of the point sources. We show theoretically and experimentally that, in the limit of a zero axial separation, our scheme allows for reaching the quantum Cram\'er-Rao lower bound and thus can be considered as one of the optimal measurement methods. Unlike other superresolution schemes, this scheme does not require neither activation of fluorophores nor sophisticated stabilization control. Moreover, it is applicable to the localization of a single point source in the axial direction. Our demonstration can be useful to a variety of applications such as far-field fluorescence microscopy.
\end{abstract}

\setboolean{displaycopyright}{true}

\begin{document}

\maketitle
\section{Introduction}
Optical microscopy is one of the most important imaging modalities and has been broadly applied in various areas. One crucial metric for an optical microscope is the spatial resolution, which is typically constrained by the diffraction limit, and the Rayleigh criterion is proposed as the resolution limit of an incoherent imaging system \cite{rayleigh1879xxxi,goodman2008introduction,born2013principles}. In recent decades, various methods have been proposed to surpass the diffraction limit. In fluorescence microscopy, a widely used approach is to activate each fluorescence molecule individually, and therefore the overlap between neighboring molecules is avoided and the localization precision can be improved to tens of nanometers \cite{betzig2006imaging, rust2006sub,hell1994breaking}. This technique usually requires specially prepared samples, and the reconstruction of an image can take a long time due to the sophisticated activation and detection of individual fluorophores. Another superresolution technique is based on decomposing the optical field into the linear prolate spheroidal functions, i.e. the eigenfunctions of aperture in a coherent imaging system \cite{beskrovnyy2005quantum,kolobov2007quantum,kolobov2000quantum}. It is shown that the ultimate limit of resolution of a coherent imaging system is not determined by diffraction but by the signal-to-noise ratio of the measurement. Therefore, a sufficiently large number of photons are needed to enable the superresolution. In addition, this technique, including other approaches that require nonclassical light sources \cite{rozema2014scalable,treps2002surpassing,taylor2014subdiffraction,taylor2013biological,shin2011quantum,schwartz2013superresolution}, cannot be readily applied to incoherent superresolution imaging considered here. While many other methods have been proposed to realize axial super-localization \cite{von2017three}, such as interferometric microscope \cite{backlund2018fundamental,shtengel2009interferometric,schrader19964pi,bewersdorf2006comparison}, point spread function (PSF) engineering \cite{huang2008three, pavani2009three, jia2014isotropic,martinez1995tunable,sales1997fundamental,tamburini2006overcoming}, and multi-plane detection \cite{juette2008three,toprak2007three,dalgarno2010multiplane,abrahamsson2012fast}, these advances can be only used to precisely measure the axial location of a single point source, and it remains a challenge to determine a small axial separation when two incoherent, simultaneously emitting point sources overlap with each other.

To develop an efficient axial superresolution technique, we follow the procedure in Ref. \cite{tsang2016quantum} and formulate the estimation of axial separation in the context of quantum metrology \cite{helstrom1969quantum,giovannetti2011advances,holevo2011probabilistic,degen2017quantum,paris2009quantum}. The precision of a measurement method is typically quantified by the Fisher information, and the reciprocal of Fisher information is referred to as the Cram\'er-Rao lower bound (CRLB) and characterizes the lower bound of measurement variance for an unbiased estimator \cite{van2004detection,kay1993fundamentals}. To determine the axial location of point sources, the easiest and most commonly used approach, which we refer to as direct imaging method, is to measure the size of the PSF in the image plane and then deduce the axial positions accordingly. However, our calculation in the next section shows that the corresponding Fisher information drops to zero when the axial separation of two incoherent point sources gets close to zero. This result is not surprising because the size of PSF changes slowly when point sources are almost on focus. Nonetheless, a further calculation shows that the quantum Fisher information does not vanish for an arbitrarily small axial separation. The quantum Fisher information is the upper limit of the Fisher information that cannot be exceeded by any possible types of measurement as derived in the quantum metrology theory and can be used to quantify the maximum possible amount of information that can be obtained by a measurement. Given a non-vanishing quantum Fisher information, there should exist a type of measurement that can outperform the direct imaging method and extract the maximum possible amount of information from each photon.

In the following sections we demonstrate both theoretically and experimentally that the axial superresolution can be achieved at the single-photon level by a radial mode sorter. This radial mode sorter can losslessly project the incident photons into the radial Laguerre-Gaussian basis set. With the same amount of photons, our scheme based on the radial mode sorter can estimate the axial separation with smaller bias and standard deviation. We note that similar strategies have been studied for transverse superresolution \cite{yang2016far,paur2016achieving,nair2016far,tham2017beating,tang2016fault, vrehavcek2018optimal,vrehavcek2017multiparameter, tsang2018subdiffraction, tsang2017subdiffractionNJP, zhou2018modern}, which are based on a Hermite-Gaussian mode sorter \cite{zhou2018hermite,paur2016achieving} or mode parity decomposition \cite{nair2016far,tham2017beating}. However, we emphasize that our radial mode sorter comes from very recent advances in spatial mode sorter \cite{zhou2017sorting,zhou2018realization,zhou2018quantum} and cannot be simply realized by mode parity decomposition. Furthermore, while homodyne or heterodyne detection \cite{yang2016far,taylor2014subdiffraction} provides an easier way to implement the spatial mode projective measurement, the shot noise from reference beam fundamentally deteriorates the signal-to-noise ratio, and it has been shown that homodyne or heterodyne detection cannot even outperform direct imaging method when the available photon number is small \cite{yang2017fisher}. Moreover, the mode sorter does not require any active components, such as the local oscillator in heterodyne detection, and thus is more favorable in an experiment.

\section{Theory}

\begin{figure}[t]
\includegraphics[width=0.9\linewidth]{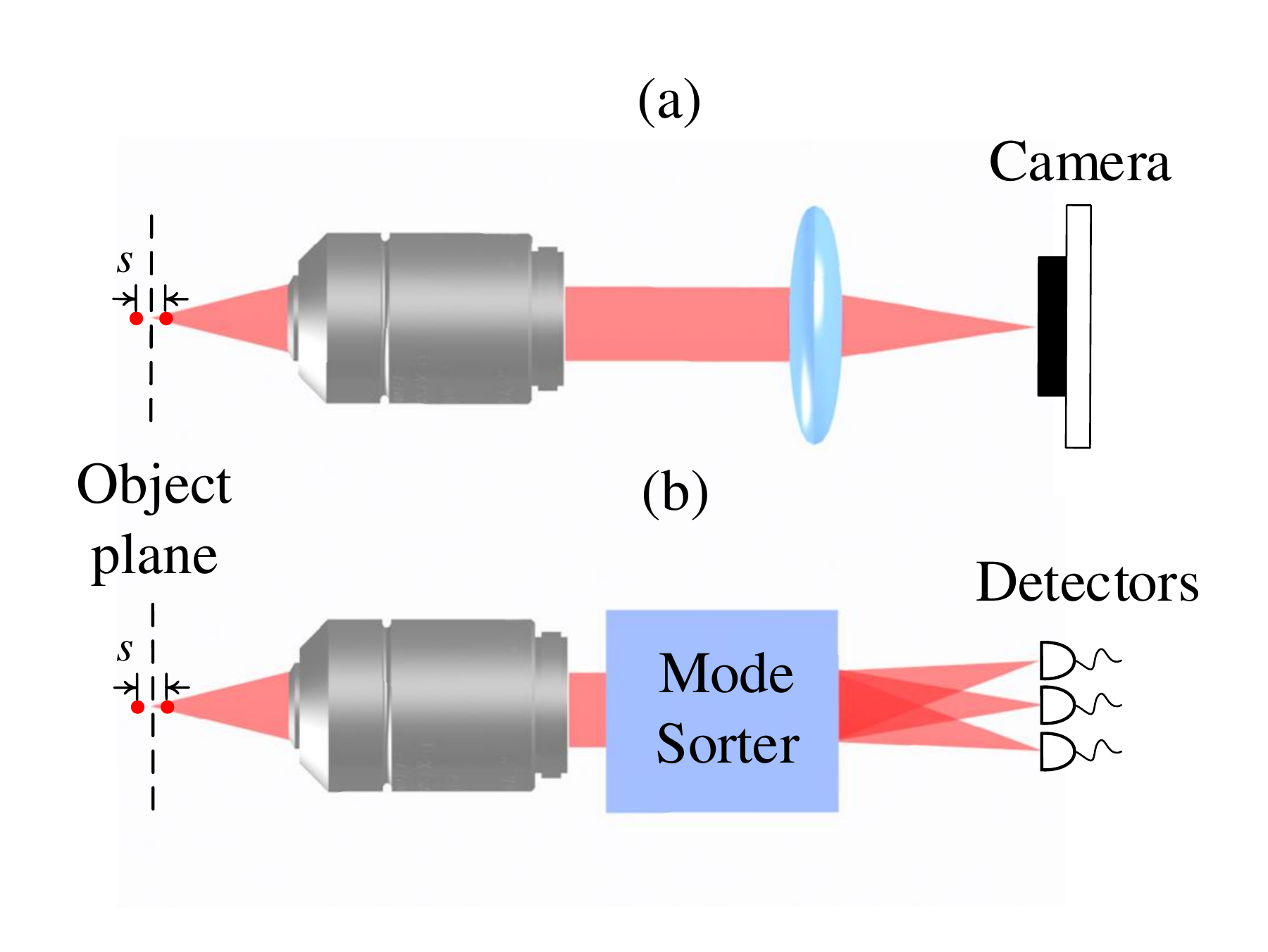}
\caption{Conceptual diagram for (a) direct imaging and (b) sorter-based measurement. A spatial mode sorter can direct different spatial mode components to different locations to perform spatial mode demultiplexing.}
\label{fig:setup}
\end{figure}

The conceptual diagram for direct imaging method and sorter-based measurement is shown in Fig.~\ref{fig:setup}. The direct imaging method employs an objective to collect photons and then use a tube lens to form an image of the object as shown in Fig.~\ref{fig:setup}(a). Alternatively, one can detect the optical field in a complete and orthonormal basis set as shown in Fig.~\ref{fig:setup}(b), which can be realized by a spatial mode sorter and is referred to as sorted-based measurement. In the following derivations we use the Dirac notation to represent the fields and assume a coherent state for each point source. While a semiclassical treatment is sufficient to derive these formalisms inspired by quantum metrology \cite{tsang2018subdiffraction}, the Dirac notation is convenient to denote the mixed state of the incoherent sources and makes it straightforward to extend the theory to other types of light sources such as single-photon state \cite{tham2017beating} and thermal state \cite{tsang2016quantum}. For a more tractable analysis and experiment here we assume a Gaussian PSF and the field distribution at the pupil plane for an on-axis point source is denoted by $\ket{\psi}$, where $\braket{r_0}{\psi}=\psi (r_0;z) $ and
\begin{equation}\label{Eq:pupil}
\begin{aligned}
\psi (r_0;z) =\sqrt{2/\pi}\exp(-r_0^2)  \exp(-ikz\text{NA}^2 r_0^2/2),
\end{aligned}
\end{equation}
where NA is the numerical aperture, $z$ is the axial position of the point source, $k=2\pi/\lambda$ is the wavenumber, $\lambda$ is the wavelength, $r_0$ is the normalized radial coordinate in the pupil plane, and $\ket{r_0}$ is the corresponding radial eigenstate. Here we define $r_0=r_p/(f_1 \text{NA})$, where $r_p$ is the radial coordinate in the pupil plane and $f_1$ is the objective focal length. This pupil plane field distribution can be viewed as a paraxial, Gaussian approximation to the pupil function of a hard-edged circular aperture \cite{petrov2017measurement}. For direct imaging a tube lens is used to perform a Fourier transform to the pupil function, and the intensity distribution on the image plane becomes
\begin{equation}
\begin{aligned}\label{Eq:DirectIntensity}
I(r;z) &= \frac{2}{\pi}\frac{1}{w^2(z)}\exp(-\frac{2r^2}{w^2(z)}), \\
w(z) &= \frac{M\lambda}{\pi \text{NA}} \sqrt{ 1+(z\pi \text{NA}^2/\lambda)^2},
\end{aligned}
\end{equation}
where $r$ is the radial coordinate in the image plane, $w(z)$ denotes the Gaussian beam waist width on the image plane, and $M$ is the magnification of the imaging system. By measuring the beam size we can estimate the axial position $z$. Similar to the case of transverse superresolution \cite{tsang2016quantum}, here we assume \textit{a priori} knowledge of two on-axis, equally bright incoherent point sources with the centroid located at $z=0$ plane, and the axial separation between them is $s$. The density matrix of these two point sources at the pupil plane can be written as $\rho =(\ket{\psi_1} \bra{\psi_1}+\ket{\psi_2} \bra{\psi_2})/2$, where $\braket{r_0}{\psi_1}=\psi (r_0;s/2) $ and $\braket{r_0}{\psi_2}=\psi (r_0;-s/2) $. The normalized total intensity at the image plane can be calculated as $I_s(r)\equiv \bra{r}\rho\ket{r}=[I(r;s/2)+I(r;-s/2)]/2$, where $\ket{r}$ is the radial eigenstate in the image plane, and the image plane is related to the pupil plane by the Fourier transform. For sorter-based measurement, the incident field is decomposed to an orthonormal basis set, and here we consider the radial Laguerre-Gaussian (LG) basis because we notice that the axial position only affects the radial profile of pupil function. The radial LG basis in the pupil plane can be denoted as $\ket{\text{LG}_p}$, where $\braket{r_0}{\text{LG}_p}=\text{LG}_p(r_0)$ and
\begin{equation}\label{Eq:LGpupil}
\begin{aligned}
&\text{LG}_p(r_0)=\sqrt{ 2/\pi }  \exp(-r_0^2) L_p( 2r_0^2 ),\\
\end{aligned}
\end{equation}
where $L_p(\cdot)$ is the Laguerre polynomial. While the two-dimensional LG basis involves another azimuthal index $\ell$, this radial subset with $\ell=0$ can still form a complete basis to describe the pupil function because of the rotational symmetry of the pupil function as shown in Eq.~(\ref{Eq:pupil}). Decomposing the pupil function $\psi (r_0;z)$ to this basis leads to the following radial mode distribution:
\begin{equation}
\begin{aligned}
P(p;z)&=|\braket{\psi}{\text{LG}_p}|^2 = \frac{    4z_R^2 z^{2p}    }{   (4z_R^2+z^2)^{p+1}  },
\end{aligned}
\label{Eq:sorterDistribution}
\end{equation}
where $z_R=\pi w_0^2/\lambda$ and $w_0=\lambda /\pi \text{NA}$ \cite{tsang2016quantum}. For two equally bright sources separated by $s$, the output radial mode distribution becomes $P_s(p) \equiv \bra{\text{LG}_p}\rho\ket{\text{LG}_p}=[P(p;s/2)+P(p;-s/2)]/2$. It can be noticed that for direct imaging and sorter-based measurement, the two incoherent point sources have the same response because Eq.~(\ref{Eq:DirectIntensity}) and Eq.~(\ref{Eq:sorterDistribution}) are even functions of $z$, which suggests that the analysis presented here can also be applied to single point localization.

We next compare the performance of direct imaging and the sorter-based measurement by calculating the Fisher information for both techniques. The Fisher information for direct imaging is \cite{tsang2016quantum}
\begin{equation}
\begin{aligned}
\mathcal{J_{\text{direct}}}(s) &= \int_0^{2\pi}d\phi  \int_0^{+\infty}\frac{1}{I_s(r)}\left( \frac{\partial I_s(r)}{\partial s}  \right)^2 rdr \\
&=\frac{4s^2}{(s^2+4z_R^2)^2},
\end{aligned}
\end{equation}
which is independent of the magnification $M$. The Fisher information for the sorter-based measurement is
\begin{equation}
\begin{aligned}
\mathcal{J_{\text{sorter}}}(s) = \sum_{p=0}^{\infty} \frac{1}{P_s\text (p) } \left(  \frac{\partial P_s (p) }{\partial  s}   \right)^2=\frac{4}{s^2+16z_R^2}.
\label{Eq:Csorter}
\end{aligned}
\end{equation}

\begin{figure}[t]
\includegraphics[width=\linewidth]{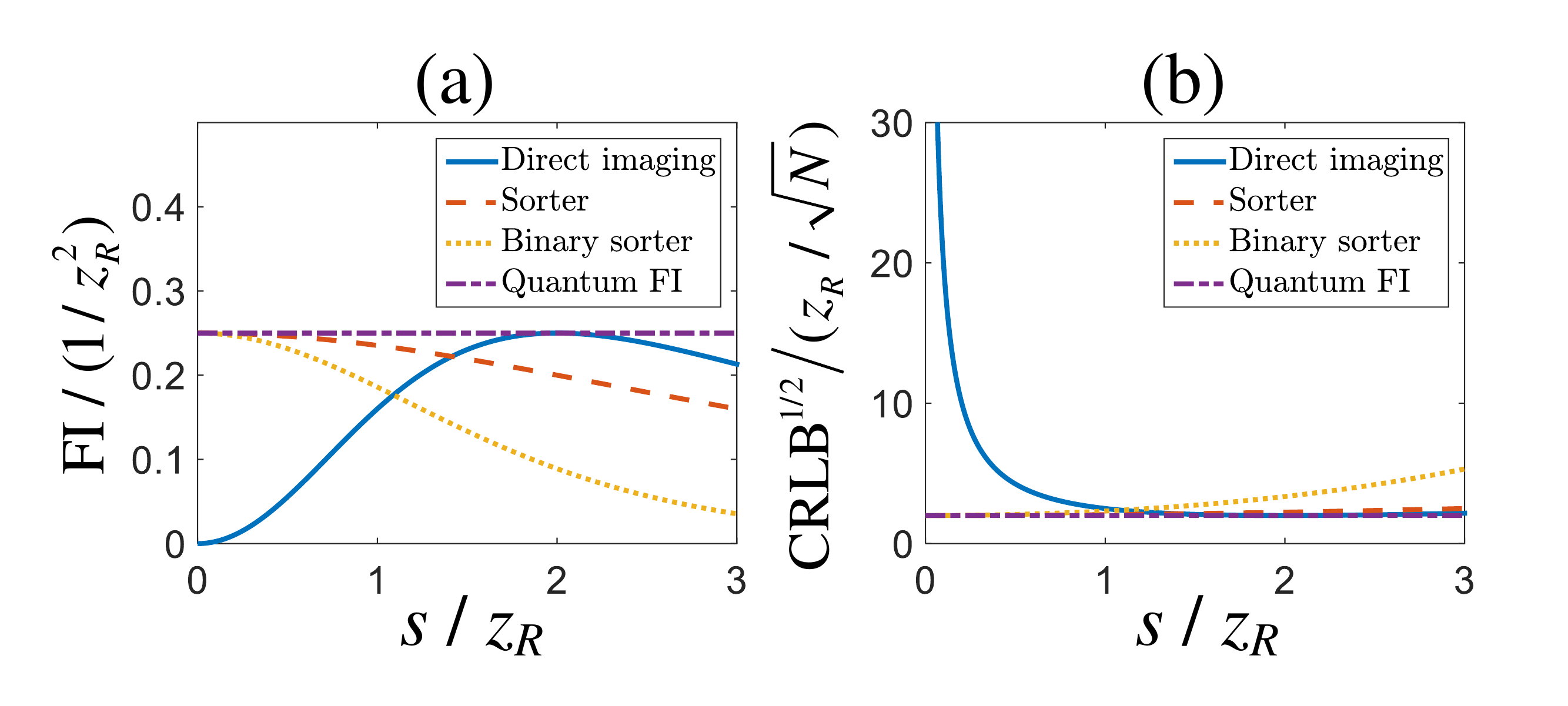}
\caption{(a) The Fisher information (FI) as a function of axial separation for different methods. The sorter and binary sorter can reach the quantum Fisher information for small separation $s$, while the Fisher information of direct imaging drops to zero. (b) The normalized square root of Cram\'er-Rao lower bound (CRLB) for different methods. $N$ is the detected photon number.}
\label{fig:FisherInfo}
\end{figure}

\begin{figure*}[t]
\center
\includegraphics[width=  0.9\textwidth]{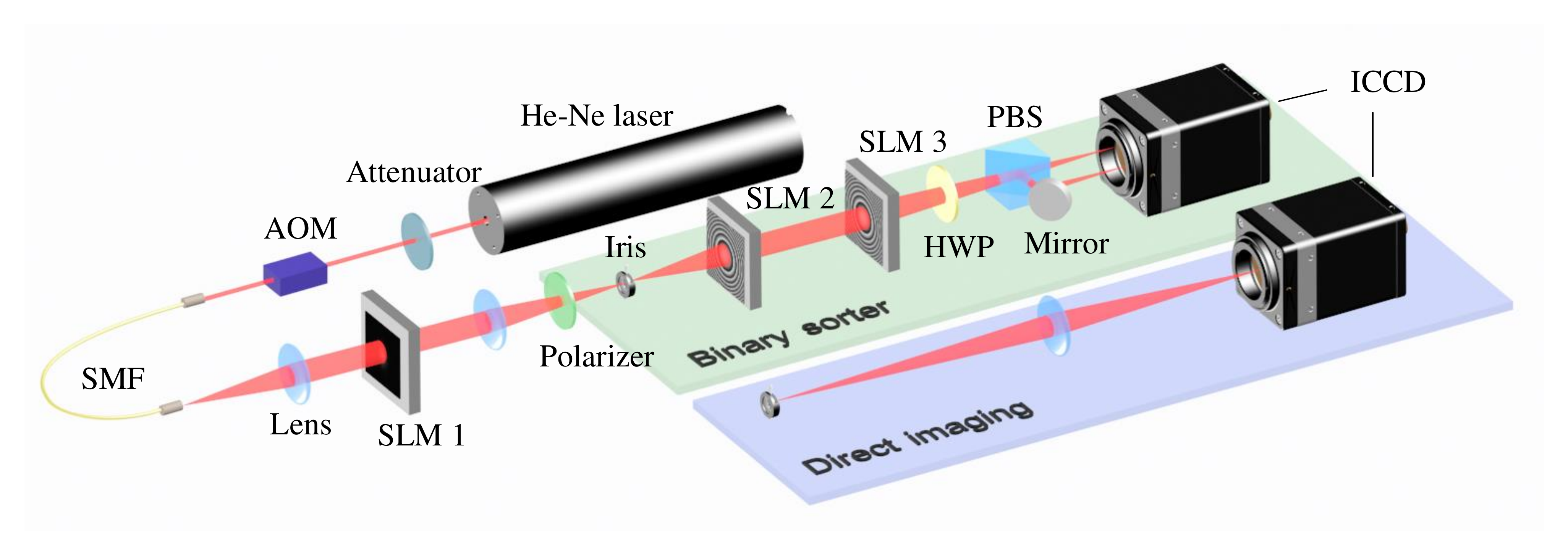}
\caption{Schematic of the experimental setup. A 633 nm He-Ne laser is attenuated and modulated by an acousto-optic modulator (AOM) to generate weak pulses. A computer-generated hologram is imprinted onto spatial light modulator (SLM 1) to generate the desired pupil function to simulate point sources. Two different methods, the binary sorter-based measurement and the direct imaging method, are used to estimate the separation $s$. In our experiment we use a flip mirror to select the measurement method.}
\label{fig:Schematic}
\end{figure*}

The quantum Fisher information, i.e. the upper bound of Fisher information of any possible measurements, can be calculated as \cite{Zhixian2018Quantum}
\begin{equation}
\begin{aligned}
\mathcal{K}_{s}=4[\braket{\partial_{s}\psi_1}{\partial_{s}\psi_1}-|\braket{\psi_1}{\partial_{s}\psi_1}|^{2}],
\end{aligned}
\end{equation}
where $\ket{\partial_s \psi_1} = \partial \ket{\psi_1} /\partial s$ and it can be readily shown that $\mathcal{K}_{s}=1/4z_R^2$. We also follow the usual way of using the symmetric logarithmic derivative to calculate the quantum Fisher information and the details are presented in Supplementary Section 1, which give the same result. The reciprocal of quantum Fisher information is the quantum CRLB which gives the lower bound of classical CRLB for any possible measurements. We notice that the sorter-based measurement can reach the quantum Fisher information when the separation goes to zero, i.e. $\mathcal{J_{\text{sorter}}}(0)= \mathcal{K}_{s} $ [see Eq.~(\ref{Eq:Csorter})], therefore it can be considered to be an optimal measurement for $s$ close to zero. However, in a realistic experiment, a mode sorter can only access a finite-dimensional Hilbert space. Therefore we follow the procedure in \cite{yang2018optimal} to construct other possible optimal measurements that can reach the quantum Fisher information in the limit of $s=0$. In Supplementary Section 2 we show that a binary radial mode sorter is sufficient to access the quantum Fisher information. A binary sorter has two output ports, and all odd-order radial modes are directed to one output port while all even-order modes are directed to another output port. Therefore, the photon probability distribution at two output ports is
\begin{equation}
\begin{aligned}
P_s^{0}(s)  &= \sum_{p=0}^{\infty} P_s(2p;s)=\frac{1}{2}+\frac{4z_R^2}{8z_R^2+s^2}, \\
P_s^{1}(s)  &= \sum_{p=0}^{\infty} P_s(2p+1;s)=\frac{1}{2}-\frac{4z_R^2}{8z_R^2+s^2}.
\end{aligned}
\end{equation}

Therefore the Fisher information for a binary sorter is
\begin{equation}
\begin{aligned}
\mathcal{J_{\text{binary}}}(s) &= \sum_{q=0}^{1} \frac{1}{P_s^q  (s) } \left(  \frac{\partial P_s^q (s) }{\partial  s}   \right)^2=\\
&=\frac{256z_R^4}{(s^2+8z_R^2)^2(s^2+16z_R^2)}.
\end{aligned}
\end{equation}
The plot of Fisher information for different methods is shown in Fig.~\ref{fig:FisherInfo}(a). It can be readily seen that the Fisher information of direct imaging begins to drop when $s$ is smaller than $2z_R$. In an incoherent imaging microscopy, the axial resolution can be expressed as $\Delta z=2\lambda/\text{NA}^2$ \cite{handbook2006JB}, which can be rewritten as $\Delta z=2\pi z_R$ with our notation. We note that the discrepancy between $2z_R$ and $2\pi z_R$ comes from our assumption of a Gaussian PSF rather than an Airy disk. However, it can be noticed that the sorter-based measurement stays nonzero and achieve the quantum Fisher information when $s$ approaches zero, which makes it possible to break the diffraction limit. To further illustrate the improvement provided by the radial mode sorter, we calculate the Fisher information of astigmatic imaging \cite{huang2008three} and the result is presented in Supplementary Section 3. It is shown that while astigmatism can enhance three-dimensional localization precision of a single point source, it cannot be used directly to resolve the axial separation between two simultaneously emitting point sources without the help of photo-switchable fluorophores.

Having analyzed the performance of each method, now we need to establish the estimator of separation. For direct imaging, it can be verified that the maximum likelihood estimator is
\begin{equation}
\begin{aligned}
\hat{w}=\sqrt{\frac{2}{N} \sum_{m=1}^{N}r_m^2}, \quad \quad \hat{s}_{\text{direct}}=2z_R \sqrt{\frac{\hat{w}^2}{w_0^2}-1}.
\label{Eq:EstDirect}
\end{aligned}
\end{equation}
where $r_m$ is the radial coordinate of $m$-th photon in the image plane and $N$ is the total detected photon number. The intuition behind this estimator is to measure the Gaussian width $w$ by detecting the radial coordinates of photons and then use the estimated Gaussian width to calculate the separation $s$. However, the simple estimator of $\hat{w}$ does not take into account any experimental imperfections such as detector noise or pixelation, and thus may not be robust in a realistic experiment. Therefore, we apply the algorithm in \cite{smith2010fast} to realize a robust, efficient Gaussian width estimator $\hat{w}$ in our experiment. For binary sorter-based measurement, the maximum likelihood estimator is
\begin{equation}
\begin{aligned}
\hat{Q}=\frac{1}{N}\sum_{q=0}^{1} q\cdot m_q=\frac{m_1}{N}, \quad \hat{s}_{\text{binary}}=2 z_R \sqrt{\frac{2}{1-2\hat{Q}} -2},
\label{Eq:EstBinary}
\end{aligned}
\end{equation}
where $m_0$ and $m_1$ are the photon numbers in the two output ports and $m_0+m_1=N$ is the total detected photon number. The intuition behind this estimator is to use the photon probability distribution at the output ports of the sorter to estimate the separation. The lower bound of the variance of an estimator for $N$ independent measurements is given by \cite{kay1993fundamentals}
\begin{equation}
\begin{aligned}
\text{Var}(\hat{s}) \geq \frac{(\partial E[\hat{s}]/\partial s)^2}{N\cdot \mathcal{J}(s)},
\label{Eq:CRLB_biased}
\end{aligned}
\end{equation}
where the right-hand side is referred to as the CRLB and $N$ is the photon number in the context of our experiment given the Poisson statistics. This formula for CRLB is also applicable to other classical photon states such as single-photon state \cite{tham2017beating} and thermal state \cite{tsang2016quantum}, and the variance that scales as $N^{-1}$ is referred to as the standard quantum limit \cite{giovannetti2004quantum,giovannetti2011advances}. For an unbiased estimator whose expectation is equal to the value of the estimated parameter, i.e. $ E[\hat{s}]=s$, this CRLB reduces to a simpler form as $\text{Var}(\hat{s}) \geq  1/ [N\cdot \mathcal{J}(s)]$, which is just the reciprocal of Fisher information as we plot in Fig.~\ref{fig:FisherInfo}(b).

\section{Experiment}
A schematic for experimental setup is shown in Fig.~\ref{fig:Schematic}. We use an attenuated laser source to illuminate the spatial light modulator (SLM) to generate the Gaussian pupil function produced by a point source. An acousto-optic modulator (AOM) is driven by a signal generator to produce 3 $\mu$s pulses, and the driving signal is also connected to an intensified charge coupled device (ICCD, PI-Max 4 1024i) for synchronization. The average detected photon number in each pulse is around 2000. We use the calibration factor provided by the manufacturer to calculate the photon number in each pixel of the camera. We emulate two incoherent sources by mixing the data for $z=\pm s/2$ that is generated by SLM separately. A computer-generated hologram is displayed on SLM~1 to generate the desired field at the first diffraction order \cite{mirhosseini2013rapid}. Each time the SLM displays the corresponding hologram to generate either $\psi (r_0;s/2)$ or $\psi (r_0;-s/2)$ to simulate a point source located at $z=s/2$ or $z=-s/2$ respectively. Since both holograms are never present at the same time, there is no coherence between the two simulated point sources. By using a long exposure time of the camera to incoherently mix the data, we effectively generate two incoherent, simultaneously emitting point sources \cite{rodenburg2014experimental}. For the Gaussian pupil function we use the parameters of $\text{NA}=0.1$ and $f_1=4$ mm. The calibration data of SLM~1 is presented in Supplementary Section 4.

To construct a binary radial mode sorter we use two polarization-sensitive SLMs (Hamamatsu X10468-02) as shown in the schematic \cite{zhou2017sorting, zhou2018quantum,zhou2018realization}. In our experiment we use two different areas on a single SLM to act as two SLMs for reduced experimental complexity. Due to the polarization sensitivity of the SLM, this binary mode sorter is designed to work for diagonally polarized light and cannot be directly used for an arbitrary polarization \cite{zhou2018quantum}, therefore we use a polarizer before the SLM to filter out undesired polarization. We note that the polarization of photons is not relevant to the theory of superresolution and thus the use of a polarization-sensitive sorter is permissible for this proof-of-principle experiment. To realize a polarization-independent sorter one can use the previously reported interferometric scheme \cite{zhou2017sorting}. A quadratic phase pattern is imprinted on SLM~2 and SLM~3 as the essential ingredient of the sorter. This quadratic phase is identical to the phase of a spherical lens with a focal length of 46.5 cm, and the separation between two SLMs is 65.8 cm. Each SLM performs a fractional Fourier transform of order $\alpha=\pi/2$ to horizontally polarized light and $\alpha=\pi/4$ to vertically polarized light respectively. One can check that even-order radial modes remain diagonally polarized and odd-order radial modes become anti-diagonally polarized after passing through both SLMs \cite{zhou2017sorting}. Through the use of a half wave-plate (HWP) and a polarizing beamsplitter (PBS) one can efficiently separate odd- and even-order radial modes to distinct output ports. More details about the principle of radial mode sorter can be found in \cite{zhou2017sorting, zhou2018quantum,zhou2018realization}. As mentioned earlier, this radial mode sorter cannot be realized by mode parity decomposition based on mirror reflection \cite{nair2016far} or a 4-$f$ system. Moreover, this radial mode sorter is in principle lossless, and the loss of our sorter mainly comes from the limited light utilization efficiency of the SLMs, which can be reduced by using other low-loss devices such as commercially available polarization directed flat lenses \cite{zhou2018realization}. In our experiment we direct the photons from different output ports to different areas of an ICCD. For direct imaging, we use a 10~cm tube lens to form the image on the ICCD detector plane. For each separation we repeat the experiment 400 times and calculate the expectation and standard deviation from the collected data based on the maximum likelihood estimators.

\section{Discussion}

\begin{figure}[t]
\includegraphics[width=\linewidth]{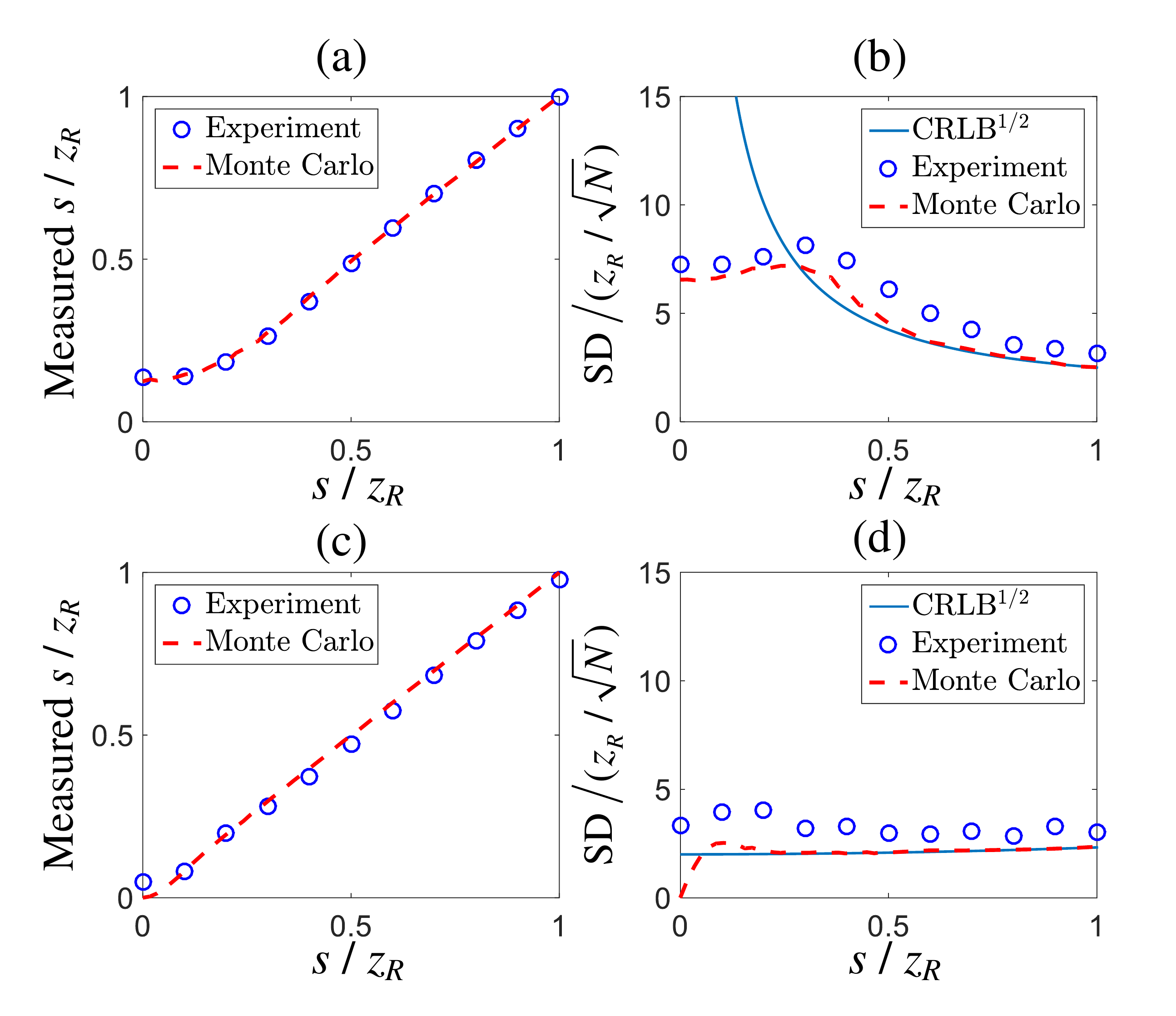}
\caption{(a) Measured separation and (b) the standard deviation (SD) of $s$ as a function of actual separation for direct imaging method. (c) Measured separation and (d) the SD of $s$ as a function of actual separation for binary sorter-based measurement. The Monte Carlo simulation results and the square root of corresponding CRLB are provided as comparisons. $N$ is the detected photon number.}
\label{fig:FisherInfoData}
\end{figure}

The measured separation and the standard deviation as a function of the real separation for different measurement methods are presented in Fig.~\ref{fig:FisherInfoData}. The average detected photon number for each measurement in our experiment is around 2000. As can be seen in Eq.~(\ref{Eq:CRLB_biased}) the loss of photons will decrease the detected photon number $N$ and thus increase the variance of measurement and deteriorate the measurement precision. However, since all devices used in our experiment are essentially phase-only elements, the loss can always be reduced to zero by using appropriate anti-reflection coatings and in our analysis we assume a detection efficiency of unity. The Monte Carlo simulation results are provided as comparisons and they agree well with the experimental data. In the simulation we set the detected photon number to be 2000, and the expectation and standard deviation of both estimators are retrieved by averaging 4000 simulations. We assume a noiseless detector with a sufficiently high spatial resolution in the simulation, and the estimators for direct imaging and binary sorter-based measurement are given by Eq.~(\ref{Eq:EstDirect}) and Eq.~(\ref{Eq:EstBinary}) respectively. One immediate observation from Fig.~\ref{fig:FisherInfoData}(a) is that the measured separation of direct imaging deviates from the real value when the real separation is close to zero. Another observation from Fig.~\ref{fig:FisherInfoData}(b) is that the measured standard deviation does not follow the CRLB and stays finite in the vicinity of $s=0$. Both observations are not due to experimental imperfections as they agree with the Monte Carlo simulation and should be attributed to the bias of the estimator \cite{tham2017beating,kay1993fundamentals}. The bias of an estimator is defined as the difference between the estimator's expectation value and the real value of the parameter being estimated. In supplementary section 5 we provide a detailed, analytical calculation of the bias of $\hat{s}_{\text{direct}}$. The expectation value of $\hat{s}_{\text{direct}}$ at $s=0$ can be well approximated as
\begin{equation}
\begin{aligned}
E[\hat{s}_{\text{direct}}]|_{s=0}\approx 0.82N^{-1/4}z_R,
\end{aligned}
\end{equation}
which is $0.123z_R$ for $N=2000$ and very close to the Monte Carlo simulation $0.124z_R$ as shown in Fig.~\ref{fig:FisherInfoData}(a). It can be noticed that this bias is on the order of $z_R$ when $N$ is small, which qualitatively agrees with the conventional axial resolution of $2\pi z_R$. A large photon number $N$ can lower the value of bias, which corresponds to the fact that higher signal-to-noise ratio can enhance the resolution of direct imaging. A simple example is the deconvolution algorithm, which can be used to obtain subdiffraction resolution as long as sufficiently high signal-to-noise ratio is available. However, the bias of direct imaging scales rather slowly with $N$ as $N^{-1/4}$, and to reduce this bias a sufficiently large $N$ is needed. While a large photon number is attainable with bright light source, in a photon-starving experiment such as fluorescence microscopy it is usually not achievable. The slope of the estimator's expectation is calculated to be
\begin{equation}
\begin{aligned}
\frac{\partial  E[\hat{s}_{\text{direct}}]}{\partial s}\bigg\rvert_{s=0} \approx \frac{0.43N^{1/4}s}{z_R}.
\end{aligned}
\end{equation}
Together with Eq.~(\ref{Eq:CRLB_biased}) it immediately follows that the CRLB becomes $\text{Var}(\hat{s}_{\text{direct}})|_{s=0}\geq 0.74z_R^2/\sqrt{N}$. In contrast to the diverging CRLB solely predicted by the reciprocal of Fisher information, the CRLB calculated here takes into account the bias and explains the non-diverging standard deviation as shown in the experiment and Monte Carlo simulation. The scaled CRLB is calculated to be $\sqrt{\text{Var}(\hat{s}_{\text{direct}})}/(z_R/\sqrt{N}) \geq 5.8$, which is close to the standard deviation in the Monte Carlo simulation result 6.6 as shown in Fig.~\ref{fig:FisherInfoData}(b). It should be noted that Eq.~(\ref{Eq:CRLB_biased}) is an inequality instead of an equality, which explains the discrepancy between 5.8 and 6.6. Furthermore, $ \partial  E[\hat{s}_{\text{direct}}]/\partial s|_{s=0} =0$ implies that the expectation value has a slope of zero when $s=0$ as can be seen in Fig.~\ref{fig:FisherInfoData}(a). Hence, despite of a finite standard deviation, it is intrinsically unrealistic to use the measured $s$ to recover the real $s$ in the vicinity of $s=0$ for the direct imaging method, and any attempt to construct an unbiased estimator will lead to a diverging standard deviation. Another observation is that the variance of the estimator for direct imaging scales as $\text{Var}(\hat{s}_{\text{direct}})|_{s=0}\propto N^{-1/2}$, therefore this estimator cannot reach the standard quantum limit when $s$ is small \cite{giovannetti2004quantum,giovannetti2011advances}.

For the sorter-based measurement, it can also be noticed that the standard deviation deviates from the reciprocal of Fisher information and drops to zero when $s$ is small. In the supplement we show that $\partial E[\hat{s}_{\text{binary}}]/\partial s|_{s=0}=0$ and thus $\text{Var}(\hat{s}_{\text{binary}})|_{s=0}\geq 0$, which explains the zero standard deviation that violates the reciprocal of Fisher information as shown in Fig.~\ref{fig:FisherInfoData}(d). It has been pointed out that this so called superefficiency only exists on a set of points with zero measure and the region of superefficiency reduces for more samples \cite{tsang2016quantum, hinkley1979theoretical}. In addition, we have also shown in the supplement that $ E[\hat{s}_{\text{binary}}] |_{s=0}=0$, which coincides with the Monte Carlo simulation and suggests that the sorter-based measurement can provide more precise, less biased measurement when $s$ is small. However, we still observe a small, nonzero separation at $s=0$ in our experiment, and the zero standard deviation is not visible either. We attribute this inconsistence to experimental imperfections including dark noise of the detector and misalignment of the sorter. At the point of $s=0$, all photons are supposed to be sorted to the output port of even-order radial modes and no photons should be detected at the other output port. Nevertheless, when we experimentally characterize our sorter we observe that 0.28\% of detected photons are routed to the wrong output port on average when $s=0$. In the data processing, we have subtracted this averaged crosstalk before estimating the separation, but the associated shot noise cannot be simply eliminated and thus leads to the experimental inconsistence as we describe above. In the supplement we quantify the effect of crosstalk on data processing and the analytical calculation shows that $ E[\hat{s}_{\text{binary}}]|_{s=0}=0.043z_R$, which is very close to the measured value of $0.049z_R$. There are several ways to further mitigate the crosstalk, such as aligning the sorter more carefully, replacing the ICCD by low-noise single-pixel detectors, and developing a more robust estimator \cite{smith2010fast}. Despite of these experimental imperfections, it is apparent that the sorter-based measurement can outperform the direct imaging, given the strong bias and higher standard deviation of direct imaging compared to that of the sorter-based measurement. In addition, the advantage of sorter-based measurement is supposed to be more obvious with a larger photon number, because the variance of direct imaging scales as $N^{-1/2}$ rather than $N^{-1}$.

\begin{figure}[t]
\centering
\includegraphics[width= \linewidth]{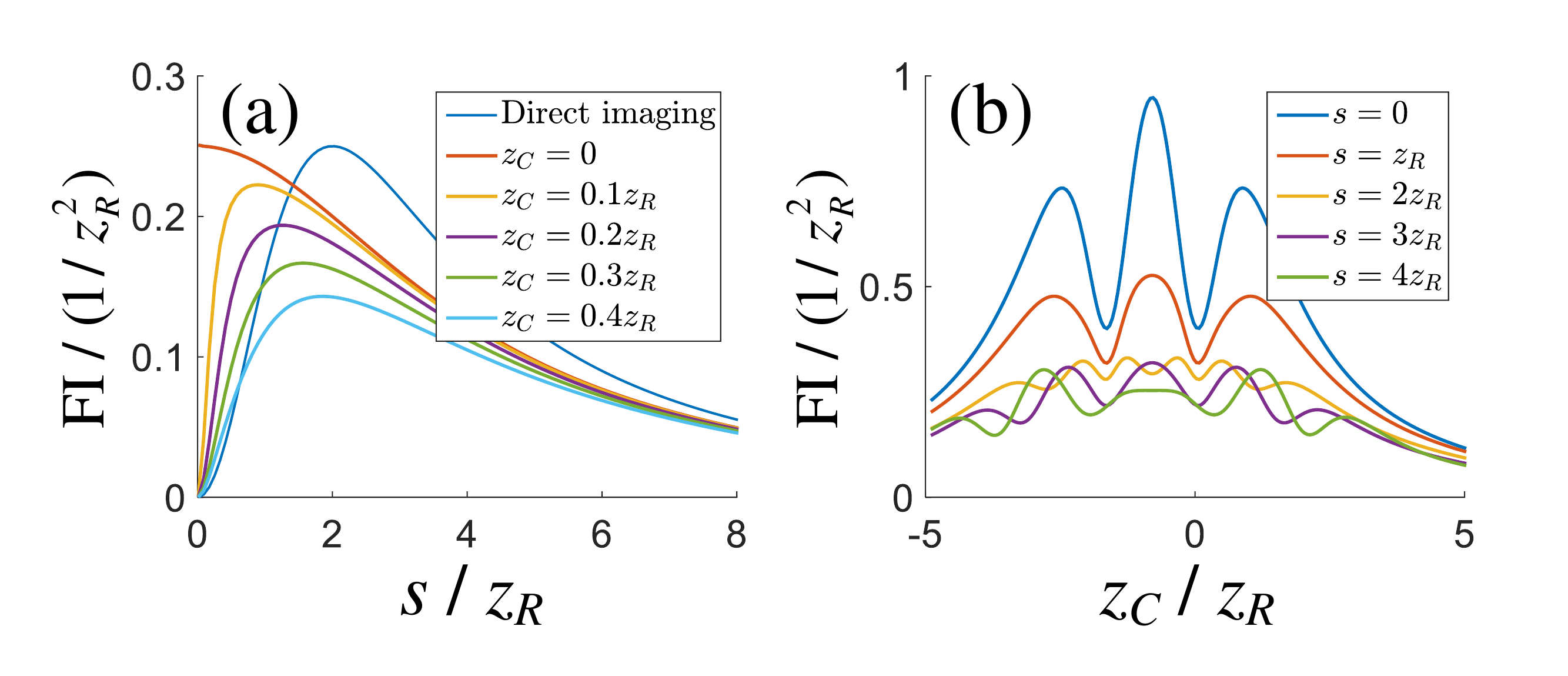}
\caption{(a) The Fisher information of separation estimation for sorter-based measurement with different centroid positions. The Fisher information for direct imaging with point source pair centroid $z_C=0$ is plotted as a reference. (b) The Fisher information of centroid estimation for astigmatic imaging with different separations.}
\label{fig:Center}
\end{figure}

\begin{figure}[b]
\centering
\includegraphics[width=0.7 \linewidth]{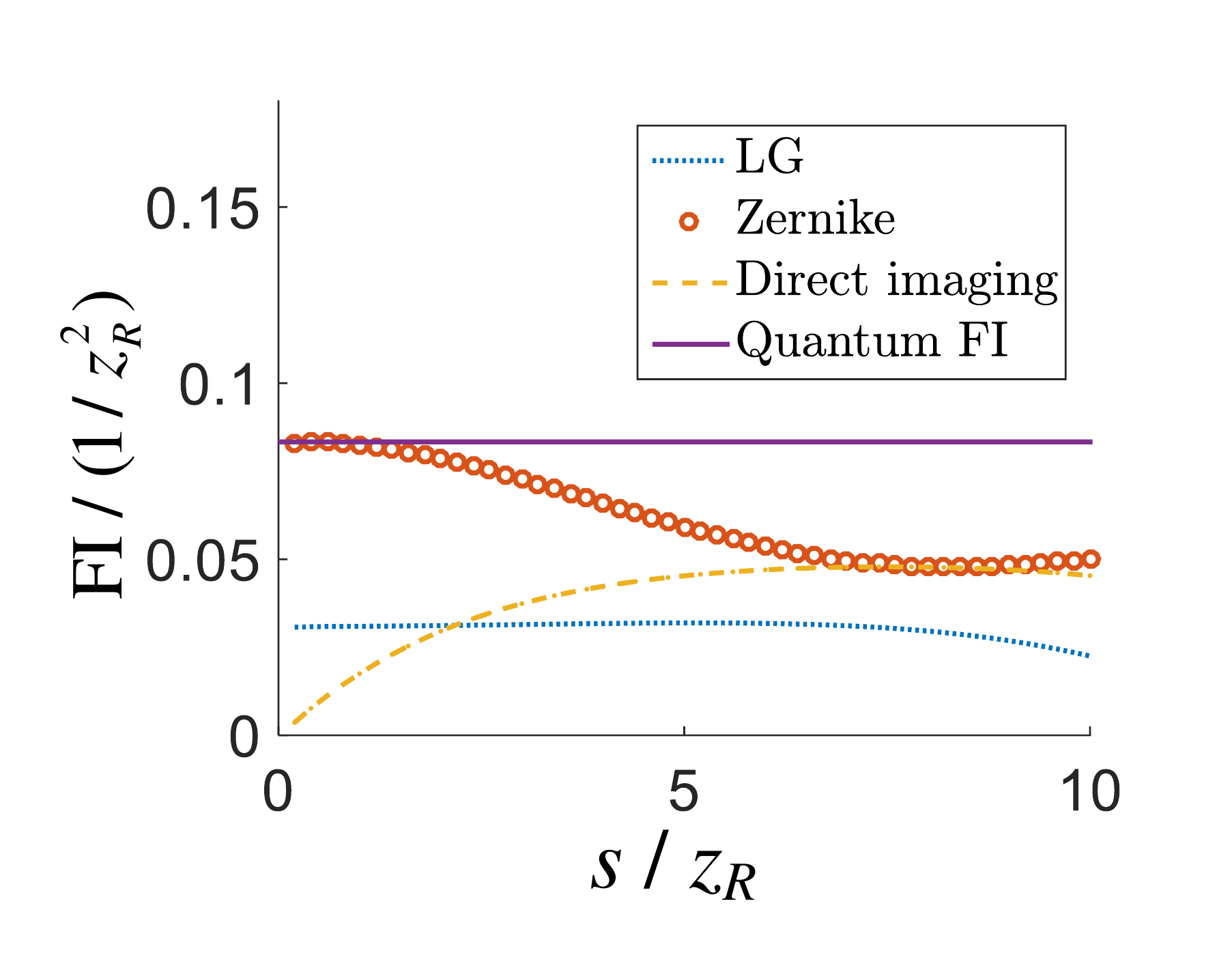}
\caption{The Fisher information of different measurements for an Airy-disk-shaped PSF model.}
\label{fig:Circ}
\end{figure}

To further provide a quantitative description of the improvement of our experiment, we compare the two methods in terms of bias and variance. For the direct imaging, if we want to reduce its bias to $0.049z_R$ that is obtained in sorter-based measurement, the photon number needs to be on the order of $10^5$ to satisfy $0.82N^{-1/4}z_R=0.049z_R$. Furthermore, from Fig.~\ref{fig:FisherInfoData} it can be seen that the measured standard deviation of direct imaging is approximately twice that of sorter-based measurement. Remember that the variance of direct imaging scales as $N^{-1/2}$, thus the standard deviation scales as $N^{-1/4}$ and 16 times more photons are needed to reduce the standard deviation of direct imaging to the level of sorter-based measurement. We note that here we are comparing the experimental data rather than noise-free theoretical predictions. The Monte Carlo simulation shows that the sorter-based measurement has zero bias and zero standard deviation at $s=0$, and thus the direct imaging needs infinite photons to beat the sorter-based measurement.

In this work we mainly focus on the superresolution of axial separation for two point sources, but we note that the theory presented above can be directly applied to the localization of axial position $z$ of a single point source as long as the separation $s$ is replaced by $s\rightarrow z/2$, which presents an alternative to the sophisticated interferometric microscopy \cite{backlund2018fundamental,shtengel2009interferometric,schrader19964pi,bewersdorf2006comparison}. In contrast to the interferometric detection scheme which requires nanometer-scale stabilization over a path length on the order of 1 m \cite{von2017three}, the common-path radial mode sorter used in our experiment is robust and no additional stabilization control is needed. We assume that the two point sources are on axis and their center position is known. In Supplementary Section 3 we analyze the effect of misaligned centroid, $i.e.$, centroid of point source pair $z_C \neq 0$. As shown in Fig.~\ref{fig:Center}(a), while the Fisher information drops in the presence of misaligned centroid, the radial mode sorter can provide improved precision for a small separation with $|z_C|<0.3z_R$. In a realistic scenario, an adaptive measurement can be used to estimate both the centroid and separation as discussed in \cite{tsang2016quantum}. However, unlike the case of transverse centroid estimation, the direct imaging does not provide sufficient Fisher information for measuring the axial centroid of point source pair. Here we notice that astigmatic imaging \cite{huang2008three} presents an effective method to overcome this difficulty. The analysis is included in Supplementary Section 3 and the Fisher information of centroid estimation for astigmatic imaging is shown in Fig.~\ref{fig:Center}(b). It can be seen that astigmatic imaging provides appreciable Fisher information over a broad range of centroid and separation. Hence, a hybrid measurement consisting of both radial mode sorter and astigmatic imaging can be a practical scheme for axial superresolution. Another assumption we make is that a Gaussian PSF is used for more tractable analysis and experiment. While the Gaussian PSF is a widely adopted approximation \cite{handbook2006JB, zhang2007gaussian}, for a high-NA imaging system a more accurate PSF model may be needed \cite{gibson1992experimental}. In this case, one can always establish a complete and orthonormal basis based on the PSF model and construct a sorter accordingly to achieve superresolution accordingly \cite{tsang2018subdiffraction}. Very recently, it has been pointed out theoretically that for the pupil function of a hard-edged aperture, the optimal measurement basis turns to be the Zernike basis \cite{Zhixian2018Quantum, yang2018optimal}, and we discuss other optimal measurements that are easier to implement in Supplementary Section 2. Therefore, based on our result it can be anticipated that three-dimensional superresolution can be realized as long as a Zernike mode sorter is available. In Supplementary Section 6 we calculate the Fisher information of various measurements for an Airy-disk-shaped PSF and the result is shown in Fig.~\ref{fig:Circ}. It can be seen that while the Zernike mode sorter provides the optimal measurement \cite{Zhixian2018Quantum}, the LG mode sorter as a sub-optimal measurement can still provide nonzero Fisher information at near-zero separation, outperforming the direct imaging measurement. Here we only take into account the 0th and 1st order radial modes in calculating the Fisher information of LG mode sorter and Zernike mode sorter, which should be reasonably achievable in an experiment. Moreover, given the widely used Gaussian-to-tophat laser beam shaper \cite{cheng2013compact, homburg2012gaussian}, it is possible to convert the pupil function of a hard-edged aperture to a Gaussian and then apply the radial mode sorter subsequently. Recent advances in multi-plane light conversion \cite{labroille2014efficient} also provide an alternative possible solution for building a Zernike mode sorters. Finally, despite the classical light source used in our experiment, our method can be used for other light sources such as single-photon emitters \cite{tsang2016quantum, tham2017beating}, because the quantum state of photons represents the temporal coherence of light and is generally independent from the spatial degree of freedom. Therefore, it is possible to combine the radial mode sorter and intensity correlation measurement to further increase the resolution for single-photons sources \cite{tenne2019super}.

In conclusion, we theoretically and experimentally demonstrate the axial superresolution based on a radial mode sorter. The binary radial mode sorter employed in out experiment can reach the quantum Cram\'er-Rao lower bound for an arbitrarily small axial separation. Our method makes three-dimensional superresolution imaging promising and can be potentially useful for enhancing the resolution of optical microscopes.

\section*{Funding}
U.S. Office of Naval Research (No. N000141712443 and N000141512635).

\section*{Acknowledgement}
R. W. B. acknowledges support from Canada Excellence Research Chairs Program and Natural Science and Engineering Research Council of Canada. J.~Y. and A.~N.~J. acknowledge support from National Science Foundation (No. DMR-1506081) and U.S. Army Research Office (No. W911NF-15-1-0496). We thank Mankei Tsang for helpful discussions.
\\
\\
See Supplement 1 for supporting content.

\newpage \setcounter{equation}{0} \setcounter{subsection}{0} \setcounter{section}{0}
\setcounter{figure}{0}
\renewcommand{\theequation}{S\arabic{equation}}
\renewcommand{\thefigure}{S\arabic{figure}}
\onecolumn
\section*{Supplementary Material}

\section{Derivation of the Quantum Fisher information}
Following the procedure in the supplement of Ref. \cite{Zhixian2018Quantum},
the quantum Fisher information in our case can be directly calculated as
\begin{equation}
\mathcal{K}_{s}=4[\braket{\partial_{s}\psi_{1}}{\partial_{s}\psi_{1}}-\big|\braket{\psi_{1}}{\partial_{s}\psi_{1}}\big|^{2}],\label{eq:Ks}
\end{equation}
and it can be readily verified that $\mathcal{K}_{s}=1/4z_R^2$ for the Gaussian PSF. Here we also follow an usual method to calculate the quantum Fisher information based on the symmetric logarithmic derivative. For two point sources that are located at $z=\pm s/2$ respectively, the density matrix of a single photon can be expressed as
\begin{equation}
\begin{aligned}
\rho = \frac{1}{2}(\ket{\psi_1} \bra{\psi_1}+\ket{\psi_2} \bra{\psi_2}),
\end{aligned}
\end{equation}
where $\bra{r_0}\ket{\psi_1} = \psi(r_0;s/2)$ and $\bra{r_0}\ket{\psi_1} = \psi(r_0;-s/2)$. The orthonormal bases in which the density matrix is diagonal are found to be
\begin{equation}
\begin{aligned}
\ket{e_1} &= A_1(\ket{\psi_1}+e^{-i\phi}\ket{\psi_2}), \\
\ket{e_2} &= A_2(\ket{\psi_1}-e^{-i\phi}\ket{\psi_2}), \\
A_1 &= ( 2+\frac{2}{\sqrt{1+\delta^2}})^{-1/2},\\
A_2 &= ( 2-\frac{2}{\sqrt{1+\delta^2}})^{-1/2},\\
\delta &= k\text{NA}^2s/4,\\
\phi &= \text{arctan}\delta.\\
\end{aligned}
\end{equation}
Hence the density matrix can be rewritten as
\begin{equation}
\begin{aligned}
\rho = \frac{1}{4A_1^2}\ket{e_1}\bra{e_1}+ \frac{1}{4A_2^2}\ket{e_2}\bra{e_2}.
\end{aligned}
\end{equation}

Here we also define the following orthonormal bases
\begin{equation}
\begin{aligned}
\ket{e_3}& = B_1(\ket{m_3}+e^{-i3\phi}\ket{m_4}), \\
\ket{e_4} &= B_2(\ket{m_3}-e^{-i3\phi}\ket{m_4}), \\
\ket{m_3}&=\frac{i}{c_0}\ket{\partial_s \psi_1}-c_1\ket{e_1}-c_2\ket{e_2}, \\
\ket{m_4}&=\frac{-i}{c_0}\ket{\partial_s \psi_2}-c_3\ket{e_1}-c_4\ket{e_2}, \\
\end{aligned}
\end{equation}
where $\ket{\partial_s \psi_1} = \partial \ket{\psi_1} /\partial s$, $\ket{\partial_s \psi_2} = \partial \ket{\psi_2}/\partial s$, and the coefficients are calculated to be

\begin{equation}
\begin{aligned}
B_1 &= \left( \frac{\delta^2(\sqrt{\delta^2+1}-1)}{2(\delta ^2+1)^{5/2}} \right)^{-1/2},   \\
B_2 &= \left( \frac{\delta^2(\sqrt{\delta^2+1}+1)}{2(\delta ^2+1)^{5/2}} \right)^{-1/2},   \\
c_0&=k\text{NA}^2/4,\\
c_1&=A_1\frac{-e^{i\phi}+(\delta-i)^2}{2(\delta-i)^2},\\
c_2&=A_2\frac{e^{i\phi}+(\delta-i)^2}{2(\delta-i)^2},\\
c_3&=A_1\frac{e^{i\phi}(\delta+i)^2-1}{2(\delta+i)^2},\\
c_4&=A_2\frac{-1-e^{i\phi}(\delta+i)^2}{2(\delta+i)^2}.\\
\end{aligned}
\end{equation}
Here the symmetric logarithmic derivative of density matrix is
\begin{equation}
\begin{aligned}
\mathcal{L}(\rho) = \sum_{j,k;D_j+D_k\neq 0}\frac{2}{D_j+D_k} \bra{e_j}\frac{\partial \rho}{\partial s}\ket{e_k} \ket{e_j} \bra{e_k}
\end{aligned}
\end{equation}
where $D_j=\bra{e_j}\rho\ket{e_j}$ and the matrix elements of $\mathcal{L}(\rho)$ can be calculated as
\begin{equation}
\begin{aligned}
\mathcal{L}_{11} &= c_0A_1^2\frac{-2\delta}{(\delta^2+1)^{3/2}}, \\
\mathcal{L}_{22} &= c_0A_2^2\frac{2\delta}{(\delta^2+1)^{3/2}}, \\
\mathcal{L}_{21} &=\mathcal{L}_{12}^*=\frac{-2ic_0\delta}{\delta^2+1} , \\
\mathcal{L}_{31} &= \mathcal{L}_{13}^*=\frac{2iA_1c_0}{2B_1}(e^{2i\phi}-1), \\
\mathcal{L}_{41} &= \mathcal{L}_{14}^*= \frac{-2iA_1c_0}{2B_2}(e^{2i\phi}+1), \\
\mathcal{L}_{32} &= \mathcal{L}_{23}^*= \frac{-2iA_1c_0}{2B_2}(e^{2i\phi}+1), \\
\mathcal{L}_{42} &= \mathcal{L}_{24}^*= \frac{2iA_2c_0}{2B_2}(e^{2i\phi}-1), \\
\mathcal{L}_{33} &= \mathcal{L}_{34}=\mathcal{L}_{43} = \mathcal{L}_{44} =0,\\
\end{aligned}
\end{equation}
where $\mathcal{L}_{jk}=\bra{e_j}\mathcal{L}(\rho)\ket{e_k}$. After more algebra one can calculate that the quantum Fisher information is
\begin{equation}
\begin{aligned}
\mathcal{K}_s(\rho) =  \text{Re Tr}\mathcal{L}(\rho)\mathcal{L}(\rho)\rho=c_0^2=\frac{1}{4z_R^2}.
\end{aligned}
\end{equation}

\section{Saturation of the quantum Cram\'er-Rao bound}

We are interested in the saturating the quantum Cram\'er-Rao bound in the limit of $s=0$. Similar to the proof of Corollary 1 in Ref.~\cite{yang2018optimal}, one can show that a measurement consisting of projectors $\Pi_{k}=\sum_{\alpha}\ket{\pi_{k\alpha}}\bra{\pi_{k\alpha}}$ can saturate the quantum Cram\'er-Rao bound at $s=0$ if and only if for all \textit{regular} projectors (defined as $|\braket{\psi_{1}\big|\Pi_{k}}{\psi_{1}}\big|_{s=0}|>0$) we can have
\begin{equation}
\braket{\partial_{s}^{0}\psi_{1}}{\pi_{k\alpha}}\big|_{s=0}=0,\quad \:\forall k,
\label{Eq:Saturation}
\end{equation}
where $\ket{\partial_{s}^{0}\psi_{1}}\big|_{s=0}=(\ket{\partial_{s}\psi_{1}}-\braket{\psi_{1}}{\partial_{s}\psi_{1}}\ket{\psi_{1}})\big|_{s=0}$. Such an optimal measurement can be constructed by choosing a proper trial basis $\{\pi_{k\alpha}\}$ and then follow the procedure in Ref.~\cite{yang2018optimal}: (i) Identify \textit{regular} basis vectors defined as $\braket{\psi_{1}}{\pi_{k\alpha}}\big|_{s=0}\neq0$ and \textit{null} basis vectors defined as $\braket{\psi_{1}}{\pi_{k\alpha}}\big|_{s=0}=0$. (ii) Calculate $\braket{\partial_{s}^{0}\psi_{1}}{\pi_{k\alpha}}\big|_{s=0}$ and check whether it vanishes or not. (iii) Assemble regular basis vectors satisfying Eq.~(\ref{Eq:Saturation}) as a {regular} projector $\Pi_{k}=\sum_{\alpha}\ket{\pi_{k\alpha}}\bra{\pi_{k\alpha}}$. (iv) A null basis vector $\ket{\pi_{k\alpha}}$ is \textit{flexible} if $\braket{\partial_{s}\psi_{1}}{\pi_{k\alpha}}\big|_{s=0}=0$. The rank one {flexible} projector $\Pi_{k\alpha}$ formed by a flexible basis vector can be added to any of the previous regular projectors or the following null projectors. (v) Assemble null basis vectors that are not flexible as a {null} projector $\Pi_{k}=\ket{\pi_{k\alpha}}\bra{\pi_{k\alpha}}$ defined as $\braket{\psi_{1}}{\Pi_{k}\big|\psi_{1}}\big|_{s=0}=0$. One can check that any measurement constructed from the previous procedure can satisfy Eq.~(\ref{Eq:Saturation}).

\subsection{Optimal measurement basis set for the Gaussian pupil function}

In the main text, we consider the case of the Gaussian pupil function, $\psi_{1}(r_{0})=\sqrt{2/\pi} \cdot \exp(-r_{0}^{2}) \cdot \exp(-iks\text{NA}^{2}r_{0}^{2}/4)$.
It is straightforward to find
\begin{equation}
\psi_{1}(r_{0})\big|_{s=0}=\text{LG}_{0}(r_{0}),
\end{equation}

\begin{equation}
\partial_{s}\psi_{1}(r_{0})\big|_{s=0}=-\frac{ik\text{NA}^{2}}{8}[\text{LG}_{0}(r_{0})-\text{LG}_{1}(r_{0})].\label{eq:dspsi1}
\end{equation}
We take radial Laguerre-Gaussian modes $\braket{r_{0}}{\text{LG}_{p}}=\text{LG}_{p}(r_{0})$
as a trial basis. So we find that for step (i) in the limit $s=0$, the only
regular basis vector is $\ket{\text{LG}_{0}}$ and the remaining other basis
vectors are null. (ii) It can be readily shown that
\[
\braket{\partial_{s}^{0}\psi_{1}}{\text{LG}_{0}}\big|_{s=0}=\braket{\partial_{s}\psi_{1}}{\text{LG}_{0}}\big|_{s=0}-\braket{\partial_{s}\psi_{1}}{\text{LG}_{0}}\big|_{s=0}\braket{\text{LG}_{0}}{\text{LG}_{0}}=0.
\]
(iii) We obtain
a regular projector $\Pi_{0}^{\prime}=\ket{\text{LG}_{0}}\bra{\text{LG}_{0}}$.
(iv) From Eq. (\ref{eq:dspsi1}), we find $\braket{\partial_{s}\psi_{1}}{\text{LG}_{p}}=0$ for $p\ge2$.
Thus the basis vectors with $p$ higher than one are all flexible and therefore
can freely added to any regular or null projector. (v) The only non-flexible
null basis vector is $\ket{\text{LG}_{1}}$, thus we can form a null
projector $\Pi_{1}^{\prime}=\ket{\text{LG}_{1}}\bra{\text{LG}_{1}}$.
We add the projector formed by flexible basis vectors of even and odd order to the previous regular projectors $\Pi_{0}$ and $\Pi_{1}$
respectively. Therefore the final optimal measurements can be
\begin{equation}
\Pi_{0}=\sum_{p=0}\ket{\text{LG}_{2p}}\bra{\text{LG}_{2p}}, \quad \Pi_{1}=\sum_{p=0}\ket{\text{LG}_{2p+1}}\bra{\text{LG}_{2p+1}}.
\end{equation}
Alternatively one can add the flexible projectors to $\Pi_{0}^{\prime}$
or $\Pi_{1}^{\prime}$ to give rise to optimal measurements of
\begin{equation}
\Pi_{0}=\ket{\text{LG}_{0}}\bra{\text{LG}_{0}}, \quad \Pi_{1}=1-\Pi_{0},
\end{equation}
or
\begin{equation}
\Pi_{0}=1-\Pi_{1}, \quad \Pi_{1}=\ket{\text{LG}_{1}}\bra{\text{LG}_{1}}.
\end{equation}

Any sorter that can efficiently perform the above measurements can be used to reach the quantum Fisher information when $s$ approaches 0.

\begin{figure}[t]
\center
\includegraphics[width=0.8\linewidth]{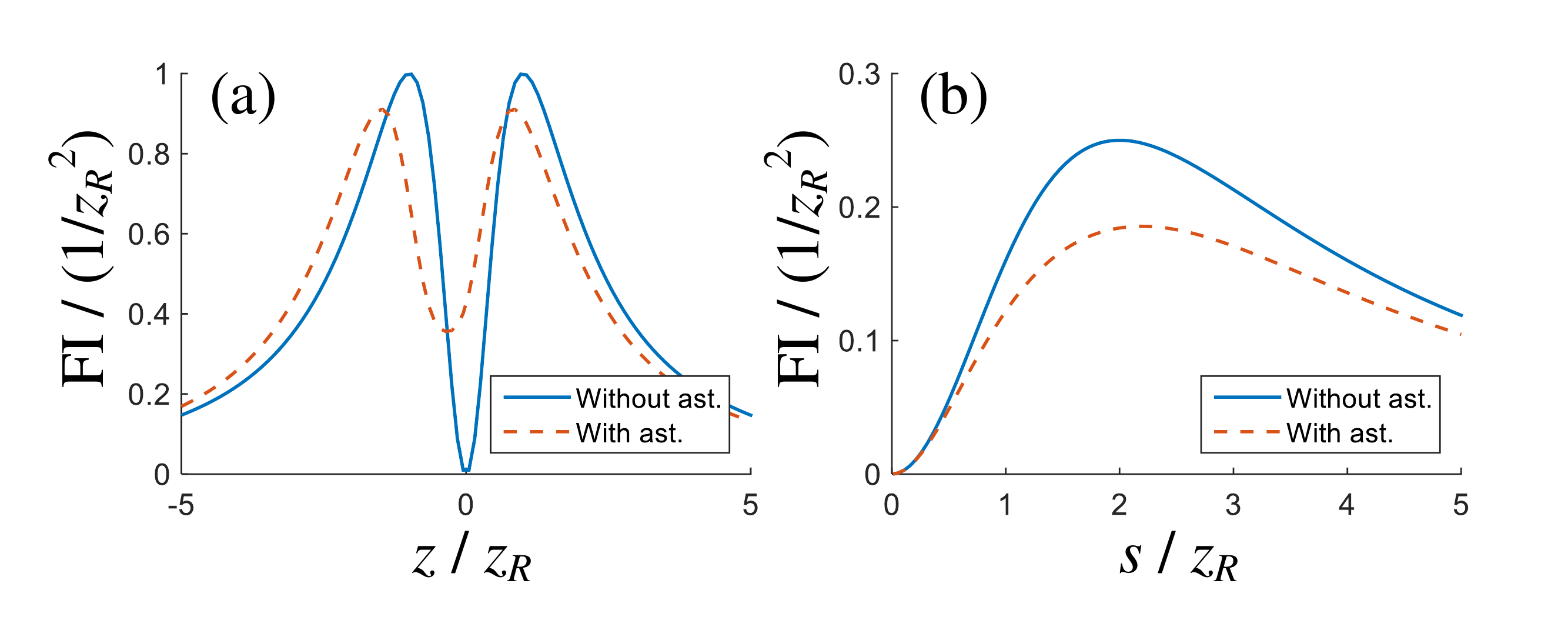}
\caption{Fisher information of (a) axial position $z$ of a single point source and (b) axial separation $s$ between a pair of point sources. The Fisher information without astigmatism (ast.) is also plotted as a reference.}
\label{fig:STORM}
\end{figure}

\subsection{Optimal measurement basis set for the pupil function of a hard-edged aperture}

The construction of optimal measurement for the Gaussian pupil
function can be analogously done for the pupil
function of hard-edged aperture. Consider the pupil function $\psi_{1}(r_{0})=\sqrt{2/\pi}\cdot \text{circ}(\sqrt{2}r_{0})\exp(-iks\text{NA}^{2}r_{0}^{2}/4)$, where $\text{circ}(r)=1$ if $0 \le r \le1$ and $\text{circ}(r)=0$ if $r >1$.
The quantum Fisher information is still expressed by Eq. (\ref{eq:Ks}).
It is straightforward to find
\begin{equation}
\psi_{1}(r_{0})\big|_{s=0}=Z_{0}(\sqrt{2}r_{0}),
\end{equation}
\begin{equation}
\partial_{s}\psi_{1}(r_{0})\big|_{s=0}=-\frac{ik\text{NA}^{2}}{16\sqrt{3}}[Z_{2}(\sqrt{2}r_{0})+\sqrt{3}Z_{0}(\sqrt{2}r_{0})],
\end{equation}
where $Z_{n}(\sqrt{2}r_{0})=\sqrt{2(n+1)/\pi}R_{n}(\sqrt{2}r_{0})\text{circ}(\sqrt{2}r_{0})$
and $R_{n}$ is the Zernike polynomial.
We take radial Zernike basis $\braket{r_{0}}{Z_{n}}=Z_{n}(\sqrt{2}r_{0})$
as the trial basis. Following the previous procedure, we find the
following optimal measurements:
\begin{equation}
\Pi_{0}=\sum_{\text{even }n}\ket{Z_{2n}}\bra{Z_{2n}}, \quad \Pi_{1}=\sum_{\text{odd }n}\ket{Z_{2n}}\bra{Z_{2n}},
\end{equation}
or
\begin{equation}
\Pi_{0}=\ket{Z_{0}}\bra{Z_{0}}, \quad \Pi_{1}=1-\Pi_{0},
\end{equation}
or
\begin{equation}
\Pi_{0}=1-\Pi_{1}, \quad \Pi_{1}=\ket{Z_{2}}\bra{Z_{2}}.
\end{equation}
Any sorter that can efficiently perform the above measurements can be used to reach the quantum Fisher information when $s$ approaches 0.

\begin{figure}[b]
\center
\includegraphics[width=0.6\linewidth]{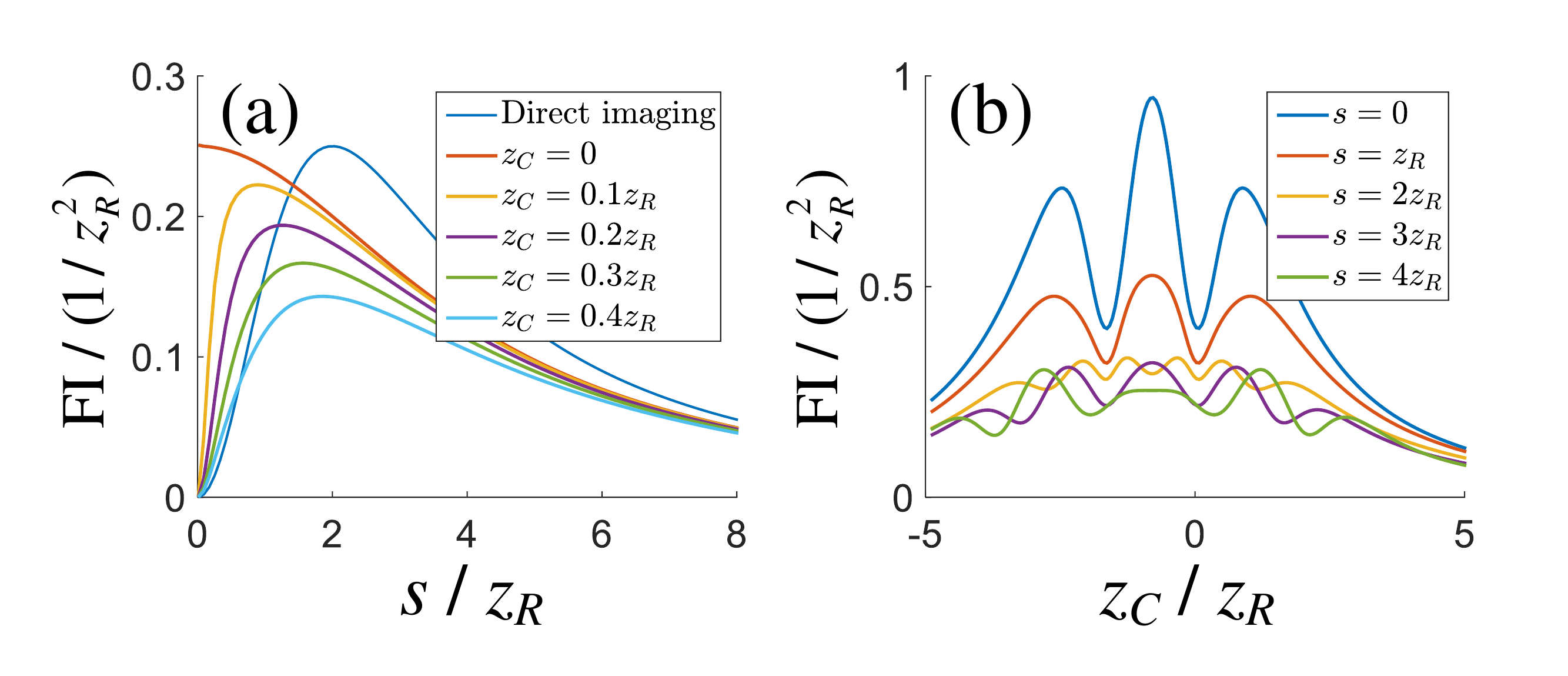}
\caption{(a) The Fisher information of separation estimation for sorter-based measurement with different centroid positions. The Fisher information for direct imaging with point source pair centroid $z_C=0$ is plotted as a reference. (b) The Fisher information of centroid estimation for astigmatic imaging with different separations.}
\label{fig:Centroid}
\end{figure}

\section{Analysis of astigmatic imaging}
\subsection{Astigmatic imaging for separation estimation}

In this section we calculate the Fisher information for astigmatic imaging. We introduce astigmatism to the Gaussian pupil function by adding the quadratic phase of a vertically oriented cylindrical lens, which can be expressed as
\begin{equation}
\begin{aligned}
\psi (x_p,y_p;z) =\sqrt{2/\pi}\frac{1}{f_1 \text{NA}}\exp(-\frac{x_p^2+y_p^2}{(f_1\text{NA})^2})  \exp(-i \frac{kz}{2f_1^2}x_p^2) \exp[-i (\frac{kz}{2f_1^2}+\frac{k}{2f_C})y_p^2],
\end{aligned}
\end{equation}
where $f_C=1.2$ m is the focal length of the astigmatic cylindrical lens and $f_1=4$ mm is the objective focal length. The corresponding intensity in the image plane can be calculated as
\begin{equation}
\begin{aligned}
I (x,y;z) =\frac{2}{\pi}\frac{1}{w_x(z)w_y(z)}\exp(-\frac{2x^2}{w_x^2(z)})\exp(-\frac{2y^2}{w_y^2(z)}),
\end{aligned}
\label{eq:IXY}
\end{equation}
where
\begin{equation}
\begin{aligned}
w_x(z)=\frac{\lambda}{\pi\text{NA}} \sqrt{1+( \pi z \text{NA}^2/\lambda)^2}, \quad w_y(z)=\frac{\lambda}{\pi\text{NA}} \sqrt{1+[ \pi (\frac{f_1^2}{f_C}+z) \text{NA}^2/\lambda]^2}.
\end{aligned}
\end{equation}
The Fisher information then can be obtained as
\begin{equation}
\begin{aligned}
\mathcal{J}_{\text{ast}}(s) = \iint_{-\infty}^{+\infty}\frac{1}{I(x,y,s/2)}(\frac{\partial I(x,y,s/2)}{\partial s})^2dxdy.
\end{aligned}
\end{equation}
The Fisher information with astigmatism is plotted as the dashed line in Fig.~\ref{fig:STORM}. As in comparison, in Fig.~\ref{fig:STORM}(a) we plot the Fisher information for single point source axial localization and in Fig.~\ref{fig:STORM}(b) the Fisher information for measurement of separation of two point sources. For axial localization of a single point source, astigmatism can help increase the Fisher information at $z \rightarrow 0$ as expected. However, for measuring the axial separation of a pair of point sources, the Fisher information again drops to zero, which contrasts to the improvement of radial mode sorter discussed in our manuscript.

\subsection{Effect of centroid misalignment and astigmatic imaging for centroid estimation}
Assume that the centroid of the point source pair is $z_C$ and the separation is $s$, so the positions of two point sources are $z_1=z_C+s/2$ and $z_1=z_C-s/2$, respectively. Therefore, the output radial mode distribution becomes $P'(p;z_C,s) \equiv \bra{\text{LG}_p}\rho\ket{\text{LG}_p}=[P(p;z_C+s/2)+P(p;z_C-s/2)]/2$. Consequently, the Fisher information of separation in the presence of centroid misalignment can be written as
\begin{equation}
\begin{aligned}
\mathcal{J'}(s) = \sum_{p=0}^{\infty} \frac{1}{P'(p;z_C,s) } \left(  \frac{\partial P'(p;z_C,s) }{\partial  s}   \right)^2,
\end{aligned}
\end{equation}
where the result is shown in Fig.\ref{fig:Centroid}(a).

As the next step, we analyze the Fisher information of centroid estimation for astigmatic imaging. The intensity at the image plane can be rewritten as
\begin{equation}
\begin{aligned}
I '(x,y;z_C,s) = \frac{1}{2}I(x,y;z_C+s/2) +  \frac{1}{2}I(x,y;z_C-s/2) ,
\end{aligned}
\end{equation}
where the definition of $I(x,y;z)$ follows Eq.~(\ref{eq:IXY}), and the Fisher information of centroid estimation is thus
\begin{equation}
\begin{aligned}
\mathcal{J}'_{\text{ast}}(z_C,s) = \iint_{-\infty}^{+\infty}\frac{1}{I(x,y;z_C,s)}(\frac{\partial I(x,y,z_C,s)}{\partial z_C})^2dxdy.
\end{aligned}
\end{equation}
Here we use the value of $f_C=0.5$ m and the result is shown in Fig.\ref{fig:Centroid}(b).

\section{SLM calibration and data processing}

\begin{figure}[t]
\center
\includegraphics[width=0.8\linewidth]{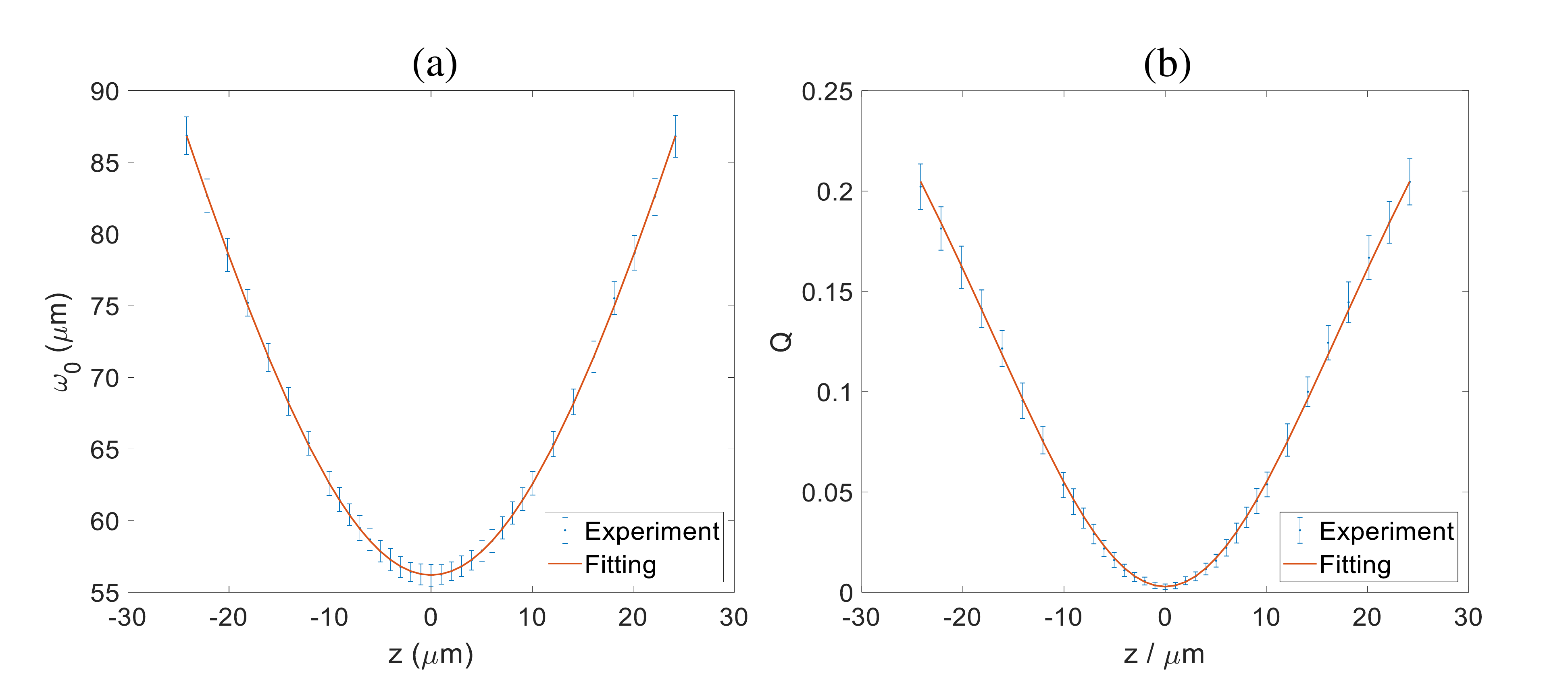}
\caption{(a) Measured Gaussian width $w_0$ as a function of $z$. (b) Measured $Q$ of binary sorter as a function of $z$.}
\label{fig:CalibrationData}
\end{figure}

In our experiment we use SLM 1 to generate Gaussian point spread function, and we calibrate SLM 1 to compensate the aberration and experimental imperfection. The pupil function we want to generate is
\begin{equation}\label{Eq:pupilS}
\begin{aligned}
\psi (r_0;z) =\sqrt{2/\pi}\exp(-r_0^2)  \exp(-ikz\text{NA}^2 r_0^2/2),
\end{aligned}
\end{equation}
where $z=\pm s/2$ and this pupil function is generated at the first diffraction order of the computer-generated hologram imprinted on the SLM. We apply Seidel abberations to the computer-generated hologram to improve mode quality and tune the parameter $z$ to fit the expected curve. For direct imaging, the measured width $w_0$ of the Gaussian PSF should increase with $z$ as expressed in Eq.~(2) in the manuscript, and the corrected experimental result is shown in Fig.~\ref{fig:CalibrationData}(a). For binary sorter-based measurement, the value of ${Q}$ calculated from the output photon numbers of the binary sorter defined in Eq. (11) can be written as a function of $z$ as
 \begin{equation}
\begin{aligned}
Q=  \frac{1}{2}\left( 1-\frac{1}{z^2/32z_R^2+1}\right).
\end{aligned}
\end{equation}
We correct the SLM to fit this curve and the calibrated data is shown in Fig.~\ref{fig:CalibrationData}(b). Due to the detector noise and misalignment of radial mode sorter, the output $Q$ has a small, nonzero value of 0.28\% at $z=0$, i.e. $Q(z=0)=0.28\%$, and we treat this experimentally measured nonzero value as a constant and subtract it before estimating the axial position. In other words, we use a new quantity $\bar{Q}=Q-0.28\%$ in the estimator to calculate $s$. However, the shot noise associated with this crosstalk cannot be simply removed by this subtraction. At $z=0$, $Q$ is in fact a random variable with an average of 0.28\%. For a specific measurement, if the measured $Q$ is lower than $0.28\%$, then $\bar{Q}$ becomes negative and the above equation results in an imaginary $z$. In our experiment, once a negative $\bar{Q}$ is obtained we force it to be zero to guarantee a real $z$. Given that $\bar{Q}$ is either zero or positive at $z=0$, the expectation of $\bar{Q}$ becomes positive rather than 0, and consequently the expectation of $\hat{z}$ becomes positive. At $z=0$ the two point sources coincide and thus this is why we measured a positive $s$ when the $s$ is zero. This treatment is also used in the estimator of direct imaging as will be analyzed in detail in the next section.

To quantify our treatment, we model the crosstalk of 0.28\% as binomial distribution. Specifically, we assume a probability of $p=0.28\%$ for a photon to be detected in the output port for odd-order radial modes. For $N$ photons, the number of photons that appear in the output port for odd-order radial modes follows the binomial distribution as
 \begin{equation}
\begin{aligned}
B(k;N,p) = \frac{N!}{k!(N-k)!} p^k (1-p)^{N-k}.
\end{aligned}
\end{equation}
Here the estimator is defined as
 \begin{equation}
\begin{aligned}
\hat{Q}=k/N  , \quad \hat{s}=2z_R\sqrt{\frac{2}{1-2\hat{Q}}-2}.
\end{aligned}
\end{equation}
The expectation of $\hat{Q}$ is $E[\hat{Q}]=p=0.28\%$. As we mentioned above, we subtract this number and use $\hat{\bar{Q}}=\hat{Q}-0.28\%$ as the new estimator to mitigate the effect of crosstalk. However, it is possible to obtain a negative $\bar{Q}$, and we force it to be zero whenever a negative value is obtained. Then the expectation of $\hat{s}$ can be calculated as
 \begin{equation}
\begin{aligned}
E[\hat{s}] = \sum_{k>Np}^{N} 2z_R B(k;H,p) \sqrt{\frac{2}{1-2\frac{k-Np}{N}}-2}.
\end{aligned}
\end{equation}
For $N=2000$ and $p=0.28\%$ the above equation can be numerically calculated to be $E[\hat{s}] =0.043z_R$, which is very close to the experimentally measured value of $0.049z_R$.

\section{Bias analysis of the estimator}
\subsection{Direct imaging method}
The intensity distribution on the image plane for the direct imaging method can be written as [see Eq.~(2) in the manuscript]
\begin{equation}
\begin{aligned}
I(r;z) &= \frac{2}{\pi}\frac{1}{w^2(z)}\exp(-\frac{2r^2}{w^2(z)}), \\
w(z) &= \frac{M\lambda}{\pi \text{NA}} \sqrt{ 1+(z\pi \text{NA}^2/\lambda)^2},
\end{aligned}
\end{equation}
which can be interpreted as the probability density function (PDF) on the image plane for single photons. The magnification $M$ is assumed to be 1 to simplify the calculation, and as we show in the manuscript this magnification does not influence the Fisher information. Since the point sources are located at $z=\pm s/2$, in the following we rewrite $w(z)$ as $w(s)=w(z=s/2)$. The maximum likelihood estimator of direct imaging for $N$ measurements is
\begin{equation}
\begin{aligned}
\hat{s}_{\text{direct}}=2z_R \sqrt{\frac{2}{Nw_0^2} \sum_{m=1}^{N}r_m^2-1},
\label{Eq:EstiDirect}
\end{aligned}
\end{equation}
and our goal here is to analyze the bias of this estimator at the limit of $s\rightarrow 0$. The CRLB of an estimator is \cite{kay1993fundamentals}
\begin{equation}
\begin{aligned}
\text{Var}(\hat{s}) \geq \frac{(\partial E[\hat{s}]/\partial s)^2}{N\cdot I(s)},
\end{aligned}
\end{equation}
where $\text{Var}(\hat{s})$is the variance of the estimator, $E[\hat{s}]$ is the expectation of estimator, and $I(s)=4s^2/(s^2+4z_R^2)^2$ is the Fisher information. For an unbiased estimator, i.e. $E[\hat{s}]=s$ and thus $\partial E[\hat{s}]/\partial s=1$, the CRLB is simply given by the inverse of Fisher information. In the manuscript we show that the Fisher information of direct imaging vanishes when $s \rightarrow 0$, but the variance of direct imaging in both experiment and simulation does not diverge to infinity. This is because of the bias of the estimator as we analyze in the following. In the polar coordinate, the PDF becomes
\begin{equation}
\begin{aligned}
f(r) = \frac{4}{w^2(s)}\exp(-\frac{2r^2}{w^2(s)}), \quad r \geq 0.
\end{aligned}
\end{equation}
Since the estimator is defined in terms of $r^2$, here we define $\bar{r}=r^2$ and the PDF of $\bar{r}$ can be calculated as
\begin{equation}
\begin{aligned}
\bar{f}(\bar{r}) = f(r(\bar{r})) \frac{dr}{d\bar{r}}=\frac{2}{w^2(s)}\exp(-\frac{2\bar{r}}{w^2(s)}), \quad r \geq 0,
\end{aligned}
\end{equation}
where $r(\bar{r})=\sqrt{\bar{r}}$. For $N$ independent measurements ($r_1, r_2,\cdots,r_m$), the PDF of the sum of random variables is the convolution of each PDF. Therefore, the PDF of $x=\sum_{m=1}^{N}r_m^2$ can be expressed as $\bar{f}(\bar{r})  \otimes \bar{f}(\bar{r}) \otimes \cdots \otimes \bar{f}(\bar{r})$, where $\otimes$ denotes the operation of convolution. This PDF turns to be the Erlang distribution, which is a special case of the gamma distribution, and can be given as
\begin{equation}
\begin{aligned}
g(x; N, k') = \frac{1}{(N-1)!}k'^N x^{N-1} \exp(-k'x), \quad x \geq 0,
\end{aligned}
\end{equation}
where $k'=2/w^2(s)$. Therefore, the PDF of $\bar{x}=2x/Nw_0^2-1$, which is the term under the square root of Eq.~(\ref{Eq:EstiDirect}), can be readily obtained as
\begin{equation}
\begin{aligned}
P(\bar{x})= g(\bar{x}+1; N, k), \quad \bar{x} \geq -1,
\end{aligned}
\end{equation}
where $k=Nw_0^2k'/2=N/[1+(s\pi \text{NA}^2/2\lambda)^2]$. It can be noticed that the estimator, which can be written as $\hat{s}_{\text{direct}}=2z_R \sqrt{\bar{x}}$, become ill-posed when $\bar{x}<0$. The intuition behind this behavior is that the estimator calculates the separation $s$ based on the measured $w(s)$. By definition we know that $w(s) \geq w_0$, but in an experiment it is possible to get a value of $w(s)$ that is even less than $w_0$, especially when the available photon number is small. One way to proceed is to force $\bar{x}=0$ whenever a negative $\bar{x}=0$ is measured, and this is how we perform our experiment as well as the Monte Carlo simulation. Hence, the integration region $-1 \leq \bar{x} \leq 0$ can be ignored because $\sqrt{\bar{x}}$ is forced to be 0, and consequently the expectation becomes
\begin{equation}
\begin{aligned}
E[\hat{s}_{\text{direct}}] = E[2z_R \sqrt{\bar{x}} ] &=2z_R \int_{0}^{\infty}\sqrt{\bar{x}} g(\bar{x}+1; N, k) d\bar{x}\\
&=2z_R \int_{1}^{\infty}\sqrt{\bar{x}-1} g(\bar{x}; N, k) d\bar{x}.
\end{aligned}
\end{equation}
When $s$ is small, we can use the Taylor expansion of $k$ as $k\approx N(1-c_0s^2)$ with $c_0=\pi^2\text{NA}^4/4\lambda^2=1/4z_R^2$. In addition, when $N$ is relatively large, such as 2000 used in our experiment, we can use the Stirling approximation $N!\approx \sqrt{2\pi}N^{N+1/2}e^{-N}$ and thus $g(\bar{x}; N, k)$ can be approximated as
\begin{equation}
\begin{aligned}
g(\bar{x}; N, k) \approx g(\bar{x}; N, N(1-c_0s^2))\approx \sqrt{\frac{N}{2\pi}} (1-c_0s^2)^N x^{N-1} e^{(1-x)N} e^{Nxc_0s^2}.
\end{aligned}
\end{equation}
Using $e^{Nxc_0s^2} \approx 1+Nxc_0s^2$ and $(1-c_0s^2)^N \approx 1-Nc_0s^2$, we have $e^{Nxc_0s^2} (1-c_0s^2)^N \approx 1+N(x-1)c_0s^2$ and thus
\begin{equation}
\begin{aligned}
g(\bar{x}; N, k) &\approx \sqrt{\frac{N}{2\pi}} x^{N-1} e^{(1-x)N}  [1+N(x-1)c_0s^2],\\
\frac{\partial g(\bar{x}; N, k) }{\partial s} & \approx \sqrt{\frac{2N}{\pi}} x^{N-1} e^{(1-x)N}  N(x-1)c_0s,
\end{aligned}
\end{equation}
therefore
\begin{equation}
\begin{aligned}
E[\hat{s}_{\text{direct}}]\big\rvert_{s=0} = \sqrt{\frac{2N}{\pi}}z_R \int_{1}^{\infty} \sqrt{\bar{x}-1}\bar{x}^{N-1}
 e^{-(x-1)N} .
\end{aligned}
\end{equation}
For CRLB, we have
\begin{equation}
\begin{aligned}
 \frac{\partial  E[\hat{s}_{\text{direct}}]}{\partial s}\bigg\rvert_{s=0}  &=2z_R \int_{1}^{\infty}\sqrt{\bar{x}-1} \frac{\partial g(\bar{x}; N, k)}{\partial s} d\bar{x}\\
&= \sqrt{\frac{N}{2\pi}}\frac{Ns }{z_R}   \int_{1}^{\infty} (\bar{x}-1)^{3/2} \bar{x}^{N-1} e^{-(\bar{x}-1)N} d\bar{x}.
\label{Eq:partialE}
\end{aligned}
\end{equation}
In the manuscript we evaluate scaled standard deviation as $\text{Var}(\hat{s})^{1/2}/(z_R/\sqrt{N})$, which can be expressed as
\begin{equation}
\begin{aligned}
\frac{\text{Var}(\hat{s})^{1/2}}{z_R/\sqrt{N}}\bigg\rvert_{s=0} &\geq  \frac{\partial E[\hat{s}_{\text{direct}}] / \partial s}{z_R I(s)} = \frac{2z_R}{s} \frac{\partial  E[\hat{s}_{\text{direct}}]}{\partial s} \\
&=\sqrt{\frac{2N}{\pi}}N\int_{1}^{\infty} (\bar{x}-1)^{3/2} \bar{x}^{N-1} e^{-(\bar{x}-1)N} d\bar{x}.
\end{aligned}
\end{equation}
The integral of above equations can be analytically calculated by the following approximations. Eq.~(\ref{Eq:partialE}) can be rewritten as
\begin{equation}
\begin{aligned}
 \frac{\partial  E[\hat{s}_{\text{direct}}]}{\partial s}\bigg\rvert_{s=0}  = \frac{s}{\sqrt{2\pi}z_R} N^{3/2}  \int_{1}^{\infty} (\bar{x}-1)^{3/2} \bar{x}^{N-1} e^{-(\bar{x}-1)N} d\bar{x}.
\end{aligned}
\end{equation}
Then we set $\bar{x}=1+t/\sqrt{N}$ and therefore the integral becomes
\begin{equation}
\begin{aligned}
 \frac{\partial  E[\hat{s}_{\text{direct}}]}{\partial s}\bigg\rvert_{s=0}  	= \frac{s}{\sqrt{2\pi}z_R}  N^{1/4}  \int_{0}^{\infty} t^{3/2}(1+\frac{t}{\sqrt{N}})^{N-1} e^{-\sqrt{N}t} dt.
\end{aligned}
\end{equation}
Now we can use the following approximation
\begin{equation}
\begin{aligned}
(1+\frac{t}{\sqrt{N}})^{N-1} &= \exp((N-1)\ln (1+\frac{t}{\sqrt{N}}))  \\
&=\exp((N-1)    (\frac{t}{\sqrt{N}} - \frac{t^2}{2N}+\cdots) ) \\
& \approx \exp( \sqrt{N}t-\frac{t^2}{2}-\frac{t}{\sqrt{N}}+\frac{t^2}{2N} ) \\
&=\exp( \sqrt{N}t) \exp(-\frac{t^2}{2})  \exp(-\frac{t}{\sqrt{N}}+\frac{t^2}{2N} ) \\
&\approx \exp( \sqrt{N}t) \exp(-\frac{t^2}{2})  (1-\frac{t}{\sqrt{N}}+\frac{t^2}{2N}),
\end{aligned}
\end{equation}
therefore
\begin{equation}
\begin{aligned}
 \frac{\partial  E[\hat{s}_{\text{direct}}]}{\partial s}\bigg\rvert_{s=0}  &\approx \frac{s}{\sqrt{2\pi}z_R} 	\propto N^{1/4}  \int_{0}^{\infty} t^{3/2} e^{-t^2/2}(1-\frac{t}{\sqrt{N}}+\frac{t^2}{2N})dt \\
 &=  \frac{s}{\sqrt{2\pi}z_R}N^{1/4} 2^{1/4} \int_{0}^{\infty} e^{-t_1}( t_1^{1/4}-\sqrt{\frac{2}{N}}t_1^{3/4}+\frac{1}{N}t_1^{5/4})dt \\
 & \approx   \frac{s}{\sqrt{2\pi}z_R} N^{1/4} 2^{1/4} \Gamma\left(\frac{5}{4}\right)\\
 &\approx \frac{0.43N^{1/4}s}{z_R},
\end{aligned}
\end{equation}
where $t_1=t^2/2$ and $\Gamma(z)=\int_{0}^{\infty}x^{z-1}\exp(-x)dx$ is the Gamma function. Immediately we can get
\begin{equation}
\begin{aligned}
\frac{\text{Var}(\hat{s})^{1/2}}{z_R/\sqrt{N}}\bigg\rvert_{s=0} &\geq  \frac{\partial E[\hat{s}_{\text{direct}}] / \partial s}{z_R I(s)} = \frac{2z_R}{s} \frac{\partial  E[\hat{s}_{\text{direct}}]}{\partial s} \approx 0.86N^{1/4}.
\end{aligned}
\end{equation}
The right-hand side of the above inequality is calculated to be 5.8 for $N=2000$, which is close to the Monte Carlo simulation result 6.6 as shown in Fig.~4(b) in the manuscript. We note that the above equation is an inequality instead of an equation, thus our analytic result is reasonable since $5.8 < 6.6$. Another apparent observation is that the CRLB of direct imaging scales with $N$ as
\begin{equation}
\begin{aligned}
 \text{Var}(\hat{s}) \geq \frac{(\partial E[\hat{s}]/\partial s)^2}{N\cdot I(s)} \propto \frac{(N^{1/4})^2}{N}=\frac{1}{\sqrt{N}}
\end{aligned}
\end{equation}
which means that this estimator cannot reach the standard quantum limit \cite{giovannetti2004quantum}. Similar math tricks can be applied to the evaluation of $E[\hat{s}_{\text{direct}}]\big\rvert_{s=0}$ and the result is
\begin{equation}
\begin{aligned}
E[\hat{s}_{\text{direct}}]\big\rvert_{s=0}=(\frac{2}{N})^{1/4}   \frac{z_R}{\sqrt{\pi}} \Gamma\left(\frac{3}{4}\right) \approx 0.82N^{-1/4}z_R,
\end{aligned}
\end{equation}
which is $E[\hat{s}_{\text{direct}}]\big\rvert_{s=0}=0.1226z_R$ for $N=2000$. Thus this approximated result is very accurate. It can be noticed that this bias term is proportional to $N^{-1/4}$, which scales rather slowly with $N$ and a sufficiently large $N \approx  10^5$ is needed to reduce the bias to $0.05z_R$.

\subsection{Sorter-based measurement}
For the binary mode sorter used in our experiment, the probability distribution is a binomial distribution which can be described as [see Eq.~(8) in the manuscript]
\begin{equation}
\begin{aligned}
P_0(s)  &=\frac{1}{2}+\frac{4z_R^2}{8z_R^2+s^2}, \quad \quad P_1(s)  &=\frac{1}{2}-\frac{4z_R^2}{8z_R^2+s^2}.
\end{aligned}
\end{equation}
The sum of $N$ independent binomial distribution becomes another binomial distribution and the probability distribution is
\begin{equation}
\begin{aligned}
P(k;N,p) = \binom{N}{k}p^k (1-p)^{N-k},
\end{aligned}
\end{equation}
where $p=P_0(s) $. The estimator used for sorter-based measurement can be written as
\begin{equation}
\begin{aligned}
\hat{Q}=\frac{N-k}{N}, \quad \hat{s}_{\text{binary}}=2 z_R \sqrt{\frac{2}{1-2\hat{Q}} -2}.
\end{aligned}
\end{equation}
In the limit of $s\rightarrow 0$, we have $p\approx 1-s^2/16z_R^2$. Therefore
\begin{equation}
\begin{aligned}
E[ \hat{s}_{\text{binary}}] &= \sum_{k=0}^{N} 2z_R P(k;N,p) \sqrt{\frac{2}{\frac{2k}{N}-1} -2} \\
&\approx \sum_{k=0}^{N} 2z_R \frac{N!}{k!(N-k)!}(1-\frac{s^2}{16z_R^2})^k(\frac{s^2}{16z_R^2})^{N-k}\sqrt{\frac{2}{\frac{2k}{N}-1} -2}\\
&= \sum_{k=0}^{N} 2z_R \frac{N!}{k!(N-k)!}(1-\frac{s^2}{16z_R^2})^{N-k}(\frac{s^2}{16z_R^2})^{k}\sqrt{\frac{2}{\frac{2(N-k)}{N}-1} -2}.
\end{aligned}
\end{equation}
In the last step we make the substitution $k \rightarrow N-k$. When $s$ is small, we can discard all higher-order terms and only keep the term with a small $k$. We keep the terms with $k=0,1,2$ and the above equation becomes
\begin{equation}
\begin{aligned}
E[ \hat{s}_{\text{binary}}] & \approx  0+ 4z_R \sqrt{N}(1-\frac{s^2}{16z_R^2})^{N-1}(\frac{s^2}{16z_R^2}) + 2z_R \sqrt{2N}(N-1) (1-\frac{s^2}{16z_R^2})^{N-2} (\frac{s^2}{16z_R^2})^2 +\cdots\\
& \approx \frac{\sqrt{N}s^2}{4z_R}+2\sqrt{2N}z_R(N-1)(\frac{s^2}{16z_R^2})^2 +\cdots
\end{aligned}
\end{equation}
If we assume that $ Ns^2/16z_R^2 \ll 1$, then all higher-order terms can be neglected and the result can be simplified to
\begin{equation}
\begin{aligned}
E[ \hat{s}_{\text{binary}}] = \frac{\sqrt{N}s^2}{4z_R}.
\end{aligned}
\end{equation}
At the first sight this result seems problematic as the expectation value increases with $N$. This is because we make a strong assumption $Ns^2/16z_R^2 \ll 1$. Therefore when $N$ is sufficiently large, the expectation value will not increase because other terms need to be taken into account. In addition, the point we want to make with this result is that $E[ \hat{s}_{\text{binary}}]|_{s=0}=0$ as well as $\partial E[ \hat{s}_{\text{binary}}]/ \partial s|_{s=0}=0$. However, unlike the direct imaging, the Fisher information of sorter-based measurement is $I_{\text{sorter}}(s)=4/(s^2+16z_R^2)$ and is nonzero at $s=0$ as $I_{\text{sorter}}(0)=1/4z_R^2$. Therefore we have
\begin{equation}
\begin{aligned}
\text{Var}(\hat{s}_{\text{sorter}}) \big\rvert_{s=0} \geq \frac{( \partial E[\hat{s}_{\text{sorter}}] / \partial s)^2}{N \cdot I_{\text{sorter}}(s)}\bigg\rvert_{s=0} =0.
\end{aligned}
\end{equation}
Both the expectation and variance agrees with the Monte Carlo simulation.

\section{Fisher information calculation for an Airy-disk-shaped PSF}

\begin{figure}[t]
\center
\includegraphics[width=0.4\linewidth]{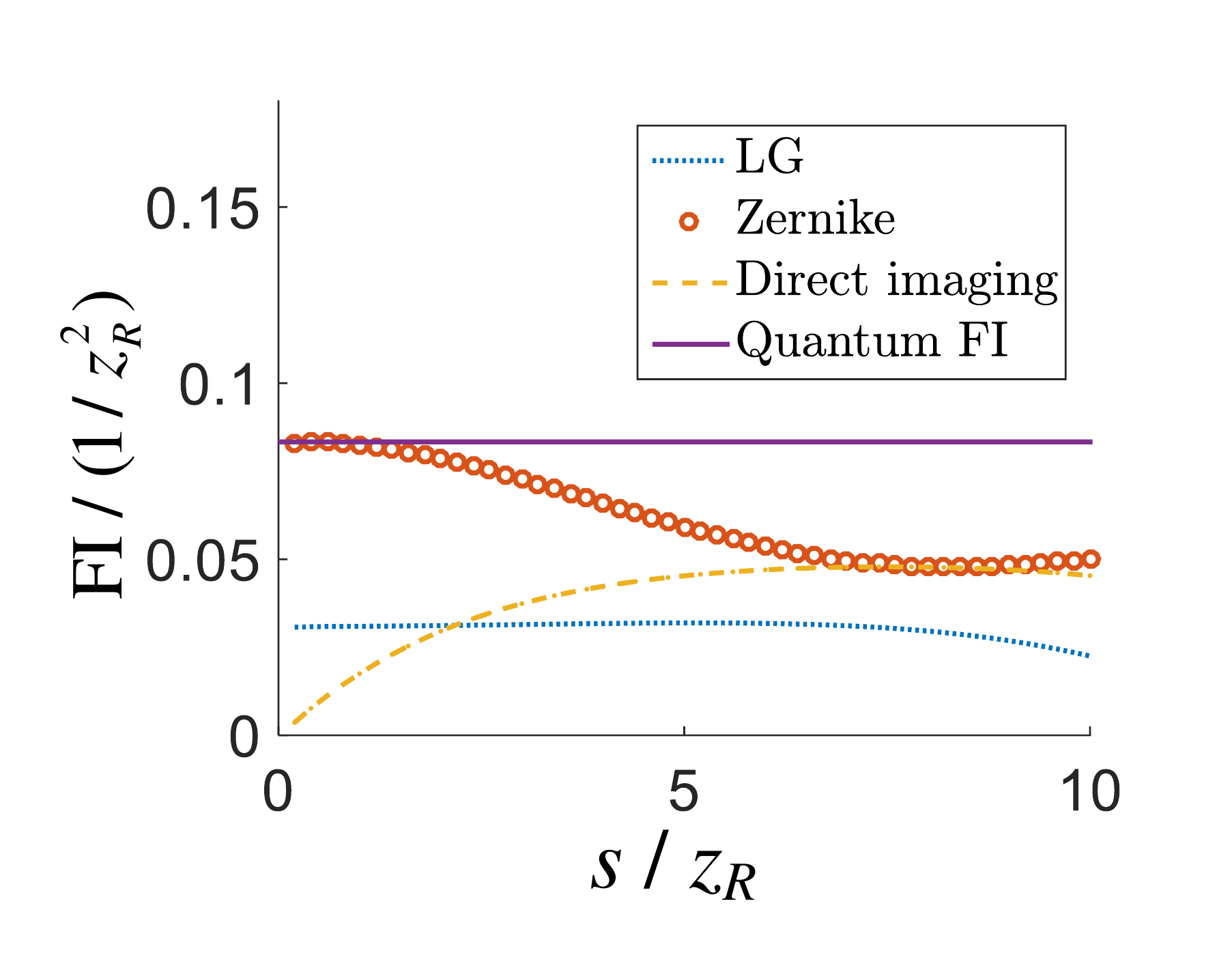}
\caption{Fisher information of different measurements for an Airy-disk-shaped PSF. }
\label{fig:CircGaussSorter}
\end{figure}

In our manuscript we mainly discuss the Gaussian PSF model which is valid only in the paraxial regime. In this section, we calculate the Fisher information for an Airy-disk-shaped PSF. The main conclusion of this section is that: (1) For an Airy-disk-shaped PSF, the Zernike mode sorter provides the optimal estimation for the axial separation of two point sources. (2) Although the Laguerre-Gaussian mode sorter is sub-optimal, it can still provide persistent nonzero Fisher information and thus outperform the direct imaging in the near-zero separation regime.

In the pupil plane, the normalized pupil function for an Airy-disk-shaped PSF can be written as
\begin{equation}
\begin{aligned}
\psi_H (r_p;z) =\frac{1}{\sqrt{\pi}(f_1 \text{NA})^2}\text{circ}( \frac{r_p}{f_1 \text{NA}}) \exp(-ikz r_p^2/2f_1^2),
\end{aligned}
\end{equation}
where $r_p$ is the radial coordinate in the pupil plane, $f_1$ is the objective focal length, and $\text{circ}(x) =1$ for $x <1$ and  $\text{circ}(x) =0$ for $x \geq 1$. The density matrix for two point sources can be expressed as $\rho_H =(\ket{\psi_H^1} \bra{\psi_H^1}+\ket{\psi_H^2} \bra{\psi_H^2})/2$, where $\braket{r_p}{\psi_H^1}=\psi_H (r_p;s/2) $ and $\braket{r_p}{\psi_H^2}=\psi_H (r_p;-s/2) $. The quantum Fisher information (QFI) can be directly calculated as \cite{Zhixian2018Quantum}
\begin{equation}
\begin{aligned}
\mathcal{K}_H =4( \langle (\partial_s\Psi)^2 \rangle  - \langle \partial_s\Psi \rangle^2  ) = \frac{\pi^2 \text{NA}^4}{12\lambda^2},
\end{aligned}
\end{equation}
where $\Psi(r_p;s) = -ks r_p^2/4f_1^2$ is the phase gradient of $\psi_H (r_p;s/2) $ and the angular bracket denotes $\langle \Phi \rangle = \int_0^{\infty} \Phi(r_p) |\psi_H (r_p;z)|^2 2\pi r_pdr_p$. It is clear that the QFI is a constant as shown in Fig.~\ref{fig:CircGaussSorter}.

The Fisher information calculation for Zernike sorter is similar to that of the radial mode sorter. Here we use the well-known Zernike polynomials as $\bra{r_p}\ket{Z_n^m} = Z_n^m(r_p/f_1\text{NA})$. The Zernike mode is normalized in such as way that $\int_0^{f_1\text{NA}} |\bra{r_p}\ket{Z_n^m}|^2 2\pi r_p dr_p =1$. Since the azimuthal part of the Zernike basis is ignored, we have $m=0$ and $n$ has to be an even number following the definition of Zernike polynomials. The Fisher information can thus be computed as
\begin{equation}
\begin{aligned}
\mathcal{J}_{\text{Zernike}}(s) = \sum_{p=0}^{p_{max}}  \left|\partial_s \bra{Z_{2p}^0}\rho_H\ket{Z_{2p}^0}\right|^2  \Big/ \bra{Z_{2p}^0}\rho_H\ket{Z_{2p}^0},
\end{aligned}
\end{equation}
where $p_{max}+1$ is the number of Zernike modes used for superresolution. In our simulation we choose $p_{max}=1$, which means that we only use two modes of lowest orders, which should be reasonably achievable in an experiment. We plot $\mathcal{J}_{\text{Zernike}}(s)$ in Fig.~~\ref{fig:CircGaussSorter} as orange circles. It can be seen that the Zernike mode sorter can reach the QFI at a near-zero separation $s$.

As a comparison, we also calculate the Fisher information for direct imaging measurement. The intensity distribution in the image plane related to the pupil function by a Fourier transform as
 \begin{equation}
\begin{aligned}
I(r,s) = \frac{1}{2}| \mathcal{F}[\psi_H(r_p,s/2)](r,s/2)|^2 + \frac{1}{2}  | \mathcal{F}[\psi_H(r_p,-s/2)](r,-s/2)|^2,
\end{aligned}
\end{equation}
where $\mathcal{F}[\cdot]$ denotes the Fourier transform with coordinate transformation from $r_p$ (in the pupil plane) to $r$ (in the image plane). Unlike the case of the Gaussian PSF, the Fourier transform of the hard-edged pupil function does not have an analytic form, so the above equation has to be computed numerically. Then the Fisher information for direct imaging measurement can be written as
 \begin{equation}
\begin{aligned}
\mathcal{J}_{\text{Direct}}(s) = \int_0^{\infty}\frac{1}{I(r,s)}\left( \frac{\partial I(r,s)}{\partial s} \right)^2 2\pi rdr,
\end{aligned}
\end{equation}
and the result is plotted as the yellow dashed line in Fig.~~\ref{fig:CircGaussSorter}. It is clear that the Fisher information drops to zero for a near-zero separation $s$.

As the next step, we calculate the Fisher information for the use of Laguerre-Gaussian (LG) mode sorter for an Airy-disk-shaped PSF instead of a Gaussian PSF. Here we use the LG basis in the pupil plane as
\begin{equation}
\begin{aligned}
\bra{r_p}\ket{\text{LG}_p} = \sqrt{\frac{2}{\pi}} \frac{1}{w_p} \exp(-\frac{r_p^2}{w_p^2})L_p\left(\frac{2r_p^2}{w_p^2}\right),
\end{aligned}
\end{equation}
where $w_p$ is the beam waist radius. Here we use the value of $w_p=f_1\text{NA}/1.121$ which is obtained by numerically maximizing the inner product $\bra{\text{LG}_0}\ket{Z_0^0}$. The intuition behind is to make $\ket{\text{LG}_0}$ be as close as possible to $\ket{Z_0^0}$. Then the Fisher information can be calculated as
\begin{equation}
\begin{aligned}
\mathcal{J}_\text{LG}(s) = \sum_{p=0}^{p_{max}}  \left|\partial_s \bra{\text{LG}_{p}}\rho_H\ket{\text{LG}_{p}}\right|^2  \Big/ \bra{\text{LG}_{p}}\rho_H\ket{\text{LG}_{p}}.
\end{aligned}
\end{equation}

The result is shown as the blue dotted line in Fig.~\ref{fig:CircGaussSorter}. It can be seen that although LG mode sorter has a lower Fisher information than the optimal Zernike mode sorter, its value at near-zero separation persists and does not drop to zero, outperforming the direct imaging.


\begin{thebibliography}{10}
\newcommand{\enquote}[1]{``#1''}

\bibitem{rayleigh1879xxxi}
L.~Rayleigh, \enquote{Xxxi. investigations in optics, with special reference to
  the spectroscope,} {\protect\JournalTitle{The London, Edinburgh, and Dublin
  Philosophical Magazine and Journal of Science}} \textbf{8}, 261--274 (1879).

\bibitem{goodman2008introduction}
J.~Goodman, \emph{Introduction to Fourier optics} (McGraw-hill, 2008).

\bibitem{born2013principles}
M.~Born and E.~Wolf, \emph{Principles of optics: electromagnetic theory of
  propagation, interference and diffraction of light} (Elsevier, 2013).

\bibitem{betzig2006imaging}
E.~Betzig, G.~H. Patterson, R.~Sougrat, O.~W. Lindwasser, S.~Olenych, J.~S.
  Bonifacino, M.~W. Davidson, J.~Lippincott-Schwartz, and H.~F. Hess,
  \enquote{Imaging intracellular fluorescent proteins at nanometer resolution,}
  {\protect\JournalTitle{Science}} \textbf{313}, 1642--1645 (2006).

\bibitem{rust2006sub}
M.~J. Rust, M.~Bates, and X.~Zhuang, \enquote{Sub-diffraction-limit imaging by
  stochastic optical reconstruction microscopy (storm),}
  {\protect\JournalTitle{Nat. Methods}} \textbf{3}, 793 (2006).

\bibitem{hell1994breaking}
S.~W. Hell and J.~Wichmann, \enquote{Breaking the diffraction resolution limit
  by stimulated emission: stimulated-emission-depletion fluorescence
  microscopy,} {\protect\JournalTitle{Opt. Lett.}} \textbf{19}, 780--782
  (1994).

\bibitem{beskrovnyy2005quantum}
V.~N. Beskrovnyy and M.~I. Kolobov, \enquote{Quantum limits of super-resolution
  in reconstruction of optical objects,} {\protect\JournalTitle{Phys. Rev. A}}
  \textbf{71}, 043802 (2005).

\bibitem{kolobov2007quantum}
M.~I. Kolobov, \emph{Quantum imaging} (Springer Science \& Business Media,
  2007).

\bibitem{kolobov2000quantum}
M.~I. Kolobov and C.~Fabre, \enquote{Quantum limits on optical resolution,}
  {\protect\JournalTitle{Phys. Rev. Lett.}} \textbf{85}, 3789 (2000).

\bibitem{rozema2014scalable}
L.~A. Rozema, J.~D. Bateman, D.~H. Mahler, R.~Okamoto, A.~Feizpour, A.~Hayat,
  and A.~M. Steinberg, \enquote{Scalable spatial superresolution using
  entangled photons,} {\protect\JournalTitle{Phys. Rev. Lett.}} \textbf{112},
  223602 (2014).

\bibitem{treps2002surpassing}
N.~Treps, U.~Andersen, B.~Buchler, P.~K. Lam, A.~Maitre, H.-A. Bachor, and
  C.~Fabre, \enquote{Surpassing the standard quantum limit for optical imaging
  using nonclassical multimode light,} {\protect\JournalTitle{Phys. Rev.
  Lett.}} \textbf{88}, 203601 (2002).

\bibitem{taylor2014subdiffraction}
M.~A. Taylor, J.~Janousek, V.~Daria, J.~Knittel, B.~Hage, H.-A. Bachor, and
  W.~P. Bowen, \enquote{Subdiffraction-limited quantum imaging within a living
  cell,} {\protect\JournalTitle{Phys. Rev. X}} \textbf{4}, 011017 (2014).

\bibitem{taylor2013biological}
M.~A. Taylor, J.~Janousek, V.~Daria, J.~Knittel, B.~Hage, H.-A. Bachor, and
  W.~P. Bowen, \enquote{Biological measurement beyond the quantum limit,}
  {\protect\JournalTitle{Nat. Photonics}} \textbf{7}, 229 (2013).

\bibitem{shin2011quantum}
H.~Shin, K.~W.~C. Chan, H.~J. Chang, and R.~W. Boyd, \enquote{Quantum spatial
  superresolution by optical centroid measurements,}
  {\protect\JournalTitle{Phys. Rev. Lett.}} \textbf{107}, 083603 (2011).

\bibitem{schwartz2013superresolution}
O.~Schwartz, J.~M. Levitt, R.~Tenne, S.~Itzhakov, Z.~Deutsch, and D.~Oron,
  \enquote{Superresolution microscopy with quantum emitters,}
  {\protect\JournalTitle{Nano Lett.}} \textbf{13}, 5832--5836 (2013).

\bibitem{von2017three}
A.~von Diezmann, Y.~Shechtman, and W.~Moerner, \enquote{Three-dimensional
  localization of single molecules for super-resolution imaging and
  single-particle tracking,} {\protect\JournalTitle{Chem. Rev.}} \textbf{117},
  7244--7275 (2017).

\bibitem{backlund2018fundamental}
M.~P. Backlund, Y.~Shechtman, and R.~L. Walsworth, \enquote{Fundamental
  precision bounds for three-dimensional optical localization microscopy with
  poisson statistics,} {\protect\JournalTitle{Phys. Rev. Lett.}} \textbf{121},
  023904 (2018).

\bibitem{shtengel2009interferometric}
G.~Shtengel, J.~A. Galbraith, C.~G. Galbraith, J.~Lippincott-Schwartz, J.~M.
  Gillette, S.~Manley, R.~Sougrat, C.~M. Waterman, P.~Kanchanawong, M.~W.
  Davidson \emph{et~al.}, \enquote{Interferometric fluorescent super-resolution
  microscopy resolves 3d cellular ultrastructure,} {\protect\JournalTitle{Proc.
  Natl. Acad. Sci. USA}} \textbf{106}, 3125--3130 (2009).

\bibitem{schrader19964pi}
M.~Schrader and S.~Hell, \enquote{4pi-confocal images with axial
  superresolution,} {\protect\JournalTitle{J. Microsc.}} \textbf{183}, 110--115
  (1996).

\bibitem{bewersdorf2006comparison}
J.~Bewersdorf, R.~Schmidt, and S.~Hell, \enquote{Comparison of i5m and
  4pi-microscopy,} {\protect\JournalTitle{J. Microsc.}} \textbf{222}, 105--117
  (2006).

\bibitem{huang2008three}
B.~Huang, W.~Wang, M.~Bates, and X.~Zhuang, \enquote{Three-dimensional
  super-resolution imaging by stochastic optical reconstruction microscopy,}
  {\protect\JournalTitle{Science}} \textbf{319}, 810--813 (2008).

\bibitem{pavani2009three}
S.~R.~P. Pavani, M.~A. Thompson, J.~S. Biteen, S.~J. Lord, N.~Liu, R.~J. Twieg,
  R.~Piestun, and W.~Moerner, \enquote{Three-dimensional, single-molecule
  fluorescence imaging beyond the diffraction limit by using a double-helix
  point spread function,} {\protect\JournalTitle{Proc. Natl. Acad. Sci. USA}}
  \textbf{106}, 2995--2999 (2009).

\bibitem{jia2014isotropic}
S.~Jia, J.~C. Vaughan, and X.~Zhuang, \enquote{Isotropic three-dimensional
  super-resolution imaging with a self-bending point spread function,}
  {\protect\JournalTitle{Nat. Photonics}} \textbf{8}, 302 (2014).

\bibitem{martinez1995tunable}
M.~Mart{\'\i}nez-Corral, P.~Andres, J.~Ojeda-Castaneda, and G.~Saavedra,
  \enquote{Tunable axial superresolution by annular binary filters. application
  to confocal microscopy,} {\protect\JournalTitle{Opt. Commun.}} \textbf{119},
  491--498 (1995).

\bibitem{sales1997fundamental}
T.~R. Sales and G.~M. Morris, \enquote{Fundamental limits of optical
  superresolution,} {\protect\JournalTitle{Opt. Lett.}} \textbf{22}, 582--584
  (1997).

\bibitem{tamburini2006overcoming}
F.~Tamburini, G.~Anzolin, G.~Umbriaco, A.~Bianchini, and C.~Barbieri,
  \enquote{Overcoming the rayleigh criterion limit with optical vortices,}
  {\protect\JournalTitle{Phys. Rev. Lett.}} \textbf{97}, 163903 (2006).

\bibitem{juette2008three}
M.~F. Juette, T.~J. Gould, M.~D. Lessard, M.~J. Mlodzianoski, B.~S. Nagpure,
  B.~T. Bennett, S.~T. Hess, and J.~Bewersdorf, \enquote{Three-dimensional
  sub--100 nm resolution fluorescence microscopy of thick samples,}
  {\protect\JournalTitle{Nat. Methods}} \textbf{5}, 527 (2008).

\bibitem{toprak2007three}
E.~Toprak, H.~Balci, B.~H. Blehm, and P.~R. Selvin, \enquote{Three-dimensional
  particle tracking via bifocal imaging,} {\protect\JournalTitle{Nano Lett.}}
  \textbf{7}, 2043--2045 (2007).

\bibitem{dalgarno2010multiplane}
P.~A. Dalgarno, H.~I. Dalgarno, A.~Putoud, R.~Lambert, L.~Paterson, D.~C.
  Logan, D.~P. Towers, R.~J. Warburton, and A.~H. Greenaway,
  \enquote{Multiplane imaging and three dimensional nanoscale particle tracking
  in biological microscopy,} {\protect\JournalTitle{Opt. Express}} \textbf{18},
  877--884 (2010).

\bibitem{abrahamsson2012fast}
S.~Abrahamsson, J.~Chen, B.~Hajj, S.~Stallinga, A.~Y. Katsov, J.~Wisniewski,
  G.~Mizuguchi, P.~Soule, F.~Mueller, C.~D. Darzacq \emph{et~al.},
  \enquote{Fast multicolor 3d imaging using aberration-corrected multifocus
  microscopy,} {\protect\JournalTitle{Nature methods}} \textbf{10}, 60 (2012).

\bibitem{tsang2016quantum}
M.~Tsang, R.~Nair, and X.-M. Lu, \enquote{Quantum theory of superresolution for
  two incoherent optical point sources,} {\protect\JournalTitle{Phys. Rev. X}}
  \textbf{6}, 031033 (2016).

\bibitem{helstrom1969quantum}
C.~W. Helstrom, \enquote{Quantum detection and estimation theory,}
  {\protect\JournalTitle{J. Stat. Phys.}} \textbf{1}, 231--252 (1969).

\bibitem{giovannetti2011advances}
V.~Giovannetti, S.~Lloyd, and L.~Maccone, \enquote{Advances in quantum
  metrology,} {\protect\JournalTitle{Nat. Photonics}} \textbf{5}, 222 (2011).

\bibitem{holevo2011probabilistic}
A.~S. Holevo, \emph{Probabilistic and statistical aspects of quantum theory},
  vol.~1 (Springer Science \& Business Media, 2011).

\bibitem{degen2017quantum}
C.~L. Degen, F.~Reinhard, and P.~Cappellaro, \enquote{Quantum sensing,}
  {\protect\JournalTitle{Rev. Mod. Phys.}} \textbf{89}, 035002 (2017).

\bibitem{paris2009quantum}
M.~G. Paris, \enquote{Quantum estimation for quantum technology,}
  {\protect\JournalTitle{Int. J. Quantum. Inform.}} \textbf{7}, 125--137
  (2009).

\bibitem{van2004detection}
H.~L. Van~Trees, \emph{Detection, estimation, and modulation theory, part I:
  detection, estimation, and linear modulation theory} (John Wiley \& Sons,
  2004).

\bibitem{kay1993fundamentals}
S.~M. Kay, \emph{Fundamentals of statistical signal processing, Volume I:
  Estimation theory (1st ed.)} (Prentice-Hall, Englewood Cliffs, 1993).

\bibitem{yang2016far}
F.~Yang, A.~Tashchilina, E.~S. Moiseev, C.~Simon, and A.~I. Lvovsky,
  \enquote{Far-field linear optical superresolution via heterodyne detection in
  a higher-order local oscillator mode,} {\protect\JournalTitle{Optica}}
  \textbf{3}, 1148--1152 (2016).

\bibitem{paur2016achieving}
M.~Pa{\'u}r, B.~Stoklasa, Z.~Hradil, L.~L. S{\'a}nchez-Soto, and J.~Rehacek,
  \enquote{Achieving the ultimate optical resolution,}
  {\protect\JournalTitle{Optica}} \textbf{3}, 1144--1147 (2016).

\bibitem{nair2016far}
R.~Nair and M.~Tsang, \enquote{Far-field superresolution of thermal
  electromagnetic sources at the quantum limit,} {\protect\JournalTitle{Phys.
  Rev. Lett.}} \textbf{117}, 190801 (2016).

\bibitem{tham2017beating}
W.-K. Tham, H.~Ferretti, and A.~M. Steinberg, \enquote{Beating rayleigh’s
  curse by imaging using phase information,} {\protect\JournalTitle{Phys. Rev.
  Lett.}} \textbf{118}, 070801 (2017).

\bibitem{tang2016fault}
Z.~S. Tang, K.~Durak, and A.~Ling, \enquote{Fault-tolerant and finite-error
  localization for point emitters within the diffraction limit,}
  {\protect\JournalTitle{Opt. Express}} \textbf{24}, 22004--22012 (2016).

\bibitem{vrehavcek2018optimal}
J.~\ifmmode \check{R}\else \v{R}\fi{}eh\'a\ifmmode~\check{c}\else \v{c}\fi{}ek,
  Z.~Hradil, D.~Koutn\'y, J.~Grover, A.~Krzic, and L.~L. S\'anchez-Soto,
  \enquote{Optimal measurements for quantum spatial superresolution,}
  {\protect\JournalTitle{Phys. Rev. A}} \textbf{98}, 012103 (2018).

\bibitem{vrehavcek2017multiparameter}
J.~\ifmmode \check{R}\else \v{R}\fi{}eha\ifmmode~\check{c}\else \v{c}\fi{}ek,
  Z.~Hradil, B.~Stoklasa, M.~Pa\'ur, J.~Grover, A.~Krzic, and L.~L.
  S\'anchez-Soto, \enquote{Multiparameter quantum metrology of incoherent point
  sources: Towards realistic superresolution,} {\protect\JournalTitle{Phys.
  Rev. A}} \textbf{96}, 062107 (2017).

\bibitem{tsang2018subdiffraction}
M.~Tsang, \enquote{Subdiffraction incoherent optical imaging via spatial-mode
  demultiplexing: Semiclassical treatment,} {\protect\JournalTitle{Phys. Rev.
  A}} \textbf{97}, 023830 (2018).

\bibitem{tsang2017subdiffractionNJP}
M.~Tsang, \enquote{Subdiffraction incoherent optical imaging via spatial-mode
  demultiplexing,} {\protect\JournalTitle{New J. Phys.}} \textbf{19}, 023054
  (2017).

\bibitem{zhou2018modern}
S.~Zhou and L.~Jiang, \enquote{A modern description of rayleigh's criterion,}
  {\protect\JournalTitle{arXiv:1801.02917}}  (2018).

\bibitem{zhou2018hermite}
Y.~Zhou, J.~Zhao, Z.~Shi, S.~M.~H. Rafsanjani, M.~Mirhosseini, Z.~Zhu, A.~E.
  Willner, and R.~W. Boyd, \enquote{Hermite-gaussian mode sorter,}
  {\protect\JournalTitle{Opt. Lett.}} \textbf{43}, 5263--5266 (2018).

\bibitem{zhou2017sorting}
Y.~Zhou, M.~Mirhosseini, D.~Fu, J.~Zhao, S.~M. Hashemi~Rafsanjani, A.~E.
  Willner, and R.~W. Boyd, \enquote{Sorting photons by radial quantum number,}
  {\protect\JournalTitle{Phys. Rev. Lett.}} \textbf{119}, 263602 (2017).

\bibitem{zhou2018realization}
D.~Fu, Y.~Zhou, R.~Qi, S.~Oliver, Y.~Wang, S.~M.~H. Rafsanjani, J.~Zhao,
  M.~Mirhosseini, Z.~Shi, P.~Zhang \emph{et~al.}, \enquote{Realization of a
  scalable laguerre--gaussian mode sorter based on a robust radial mode
  sorter,} {\protect\JournalTitle{Opt. Express}} \textbf{26}, 33057--33065
  (2018).

\bibitem{zhou2018quantum}
Y.~Zhou, M.~Mirhosseini, S.~Oliver, J.~Zhao, S.~M.~H. Rafsanjani, M.~P.~J.
  Lavery, A.~E. Willner, and R.~W. Boyd, \enquote{High-dimensional free-space
  quantum key distribution using spin, azimuthal, and radial quantum numbers,}
  {\protect\JournalTitle{arXiv:1809.09986}}  (2018).

\bibitem{yang2017fisher}
F.~Yang, R.~Nair, M.~Tsang, C.~Simon, and A.~I. Lvovsky, \enquote{Fisher
  information for far-field linear optical superresolution via homodyne or
  heterodyne detection in a higher-order local oscillator mode,}
  {\protect\JournalTitle{Phys. Rev. A}} \textbf{96}, 063829 (2017).

\bibitem{petrov2017measurement}
P.~N. Petrov, Y.~Shechtman, and W.~Moerner, \enquote{Measurement-based
  estimation of global pupil functions in 3d localization microscopy,}
  {\protect\JournalTitle{Opt. Express}} \textbf{25}, 7945--7959 (2017).

\bibitem{Zhixian2018Quantum}
Z.~Yu and S.~Prasad, \enquote{Quantum limited superresolution of an incoherent
  source pair in three dimensions,} {\protect\JournalTitle{Phys. Rev. Lett.}}
  \textbf{121}, 180504 (2018).

\bibitem{yang2018optimal}
J.~Yang, S.~Pang, Y.~Zhou, and A.~N. Jordan, \enquote{Optimal measurements for
  quantum multi-parameter estimation with general states,}
  {\protect\JournalTitle{arXiv:1806.07337}}  (2018).

\bibitem{handbook2006JB}
J.~B. Pawley, \emph{Handbook of biological confocal microscopy} (Springer
  Science \& Business Media, 2010).

\bibitem{smith2010fast}
C.~S. Smith, N.~Joseph, B.~Rieger, and K.~A. Lidke, \enquote{Fast,
  single-molecule localization that achieves theoretically minimum
  uncertainty,} {\protect\JournalTitle{Nat. Methods}} \textbf{7}, 373 (2010).

\bibitem{giovannetti2004quantum}
V.~Giovannetti, S.~Lloyd, and L.~Maccone, \enquote{Quantum-enhanced
  measurements: beating the standard quantum limit,}
  {\protect\JournalTitle{Science}} \textbf{306}, 1330--1336 (2004).

\bibitem{mirhosseini2013rapid}
M.~Mirhosseini, O.~S. Magana-Loaiza, C.~Chen, B.~Rodenburg, M.~Malik, and R.~W.
  Boyd, \enquote{Rapid generation of light beams carrying orbital angular
  momentum,} {\protect\JournalTitle{Opt. Express}} \textbf{21}, 30196--30203
  (2013).

\bibitem{rodenburg2014experimental}
B.~Rodenburg, M.~Mirhosseini, O.~S. Maga{\~n}a-Loaiza, and R.~W. Boyd,
  \enquote{Experimental generation of an optical field with arbitrary spatial
  coherence properties,} {\protect\JournalTitle{J. Opt. Soc. Am. B}}
  \textbf{31}, A51--A55 (2014).

\bibitem{hinkley1979theoretical}
D.~V. Hinkley and D.~Cox, \emph{Theoretical statistics} (Chapman and Hall/CRC,
  1979).

\bibitem{zhang2007gaussian}
B.~Zhang, J.~Zerubia, and J.-C. Olivo-Marin, \enquote{Gaussian approximations
  of fluorescence microscope point-spread function models,}
  {\protect\JournalTitle{Appl. Opt.}} \textbf{46}, 1819--1829 (2007).

\bibitem{gibson1992experimental}
S.~F. Gibson and F.~Lanni, \enquote{Experimental test of an analytical model of
  aberration in an oil-immersion objective lens used in three-dimensional light
  microscopy,} {\protect\JournalTitle{J. Opt. Soc. Am. A}} \textbf{9}, 154--166
  (1992).

\bibitem{cheng2013compact}
W.~Cheng, W.~Han, and Q.~Zhan, \enquote{Compact flattop laser beam shaper using
  vectorial vortex,} {\protect\JournalTitle{Appl. Opt.}} \textbf{52},
  4608--4612 (2013).

\bibitem{homburg2012gaussian}
O.~Homburg and T.~Mitra, \enquote{Gaussian-to-top-hat beam shaping: an overview
  of parameters, methods, and applications,} {\protect\JournalTitle{Proc.
  SPIE}} \textbf{8236}, 82360A (2012).

\bibitem{labroille2014efficient}
G.~Labroille, B.~Denolle, P.~Jian, P.~Genevaux, N.~Treps, and J.-F. Morizur,
  \enquote{Efficient and mode selective spatial mode multiplexer based on
  multi-plane light conversion,} {\protect\JournalTitle{Opt. Express}}
  \textbf{22}, 15599--15607 (2014).

\bibitem{tenne2019super}
R.~Tenne, U.~Rossman, B.~Rephael, Y.~Israel, A.~Krupinski-Ptaszek,
  R.~Lapkiewicz, Y.~Silberberg, and D.~Oron, \enquote{Super-resolution
  enhancement by quantum image scanning microscopy,}
  {\protect\JournalTitle{Nat. Photonics}} \textbf{13}, 116 (2019).

\end{thebibliography}
\end{document}